\documentclass[pre,superscriptaddress,onecolumn,floatfix,longbibliography]{revtex4-1}

\usepackage{graphics}
\usepackage{graphicx}
\usepackage{amsmath}
\usepackage{amssymb,xcolor}
\usepackage{subcaption}
\usepackage[percent]{overpic}

\definecolor{grisclair}{rgb}{0.6,0.6,0.6}

\newcommand{\beq}{\begin{equation}}
\newcommand{\ee}{\end{equation}}

\begin{document}

%\title{Experimental study of bubble rising close to a downstream-located vertical wall}
\title{Effect of a downstream vertical wall on the rise regime of an isolated bubble: an experimental study}

\author{T. González-Rubio}
\address{Depto.\ de Ingenier\'{\i}a Mec\'anica, Energ\'etica y de los Materiales and\\ 
Instituto de Computaci\'on Cient\'{\i}fica Avanzada (ICCAEx),\\
Universidad de Extremadura, E-06006 Badajoz, Spain}
\author{A. Rubio}
\address{Depto.\ de Ingenier\'{\i}a Mec\'anica, Energ\'etica y de los Materiales and\\ 
Instituto de Computaci\'on Cient\'{\i}fica Avanzada (ICCAEx),\\
Universidad de Extremadura, E-06006 Badajoz, Spain}
\author{R. Bolaños-Jiménez}
\email[Corresponding author: R. Bolaños-Jiménez ]{(rbolanos@ujaen.es)}
\address{Área de Mecánica de Fluidos, Departamento de Ingeniería Mecánica y Minera, Universidad de Jaén and\\
Andalusian Institute for Earth System Research, Campus de las Lagunillas,
23071 Jaén, Spain}
\author{E. J. Vega}
\address{Depto.\ de Ingenier\'{\i}a Mec\'anica, Energ\'etica y de los Materiales and\\ 
Instituto de Computaci\'on Cient\'{\i}fica Avanzada (ICCAEx),\\
Universidad de Extremadura, E-06006 Badajoz, Spain}

\date{July 2025}

\begin{abstract}

This work experimentally investigates deformable nitrogen bubbles rising in ultrapure water and interacting with a vertical wall, focusing on how this downstream boundary alters their dynamics, an effect critical to many real-world processes. The experiments were conducted with a fixed Morton number, $Mo = 2.64 \times 10^{-11}$, with Bond, Galilei, and Reynolds numbers in the ranges $0.08 \lesssim Bo \lesssim 0.33$, $71 \lesssim Ga \lesssim 194$, and $132 \lesssim Re \lesssim 565$, respectively. The initial dimensionless horizontal distance between the wall and the bubble centroid was systematically varied, $0.3 \lesssim L \lesssim 5$, and the bubble trajectories from two orthogonal vertical planes were captured using high-speed imaging. While the bubble rising paths were stable without the wall presence for all the cases, the results reveal that wall proximity significantly affects the rising path, depending on $Bo$ (or $Ga$) and $L$. A map with four distinct interaction regimes and their transitions is obtained: (i) Rectilinear Path (RP) at low $Bo$ and large $L$, with negligible wall influence; (ii) Migration Away (MA) at higher $Bo$ and moderate-to-large $L$, with lateral deviation from the wall; (iii) Collision and Migration Away (C+MA) at high $Bo$ and small $L$, where bubbles first collide and then migrate away; and (iv) Periodic Collisions (PC) at low $Bo$, where repeated wall impacts occur due to competing forces. These findings bridge the gap between idealised simulations and practical systems, offering high-quality data to support and refine computational models of bubble-wall interactions in industrial and environmental applications.

%This study presents an experimental investigation of deformable bubbles rising freely in an unbounded liquid until encountering a vertical wall. Our setup allows for a detailed examination of how wall interaction modifies the bubble rising dynamics. In particular, we investigate nitrogen bubbles ascending in ultrapure water ($Mo=2.64\times10^{-11}$) with Bond and Reynolds numbers in the ranges $0.08\lesssim \text{Bo}\lesssim 0.33$ and $132\lesssim  \text{Re}\lesssim 565$. An extensive experimental campaign was performed varying the horizontal distance of a vertical wall placed downstream of the gas injector, $0.3 \lesssim L \lesssim 5$, with $L$ the dimensionless gap between the wall and the injector centre. By high-speed recordings of two vertical planes, detailed data on the behaviour of bubbles during wall encounters were collected. Our results show that the wall alters the rising path depending on the value of the control parameters, $Bo$ (or Galilei) and $L$, which allowed us to obtain a map of the observed regimes with the corresponding transitions. The experimental configuration is important in real-world processes, such as aeration in tanks and multiphase reactors, where bubbles do not originate near boundaries but encounter them during ascent. The findings aim to complement recent numerical and theoretical studies, bridging the gap between controlled experiments and complex practical systems, and providing valuable data to improve modeling and understanding of bubble-boundary interactions in industrial and environmental contexts.
\end{abstract}

\maketitle

\section{Introduction}
\label{sec1}

The rise of gas bubbles in liquids is a canonical two-phase flow problem that has attracted widespread attention due to its relevance across both natural and industrial contexts. From water purification and mineral flotation to carbon capture and medical applications, bubble dynamics underpin many technologies~\citep[see, for example][and references therein]{RSMG15, WZLZC18, CM25}. Much of the foundational understanding of bubble motion has been obtained from studies of a single bubble rising in an unbounded, quiescent fluid~\citep{CGW78,BW81,D95,ME00,CMMT16, CMMT16,BFM23,RVCMLH24,BFM23,FCLHM25}. In such a configuration, the interplay of buoyancy, viscous forces, inertial effects, and surface tension determines the bubble trajectory, shape, and terminal velocity. These are typically captured by two dimensionless numbers: the Bond number ($Bo = \rho g (D^{*})^{2} / \sigma$), and the Galilei number ($Ga = \sqrt{g (D^{*})^{3}}  \rho / \mu$), with $D^{*}$ the equivalent bubble diameter, $\rho$ and $\mu$ the liquid density and viscosity, respectively, $\sigma$ the surface tension, and $g$ the gravity acceleration. The combination of these two parameters determines the resulting Reynolds number ($Re=\rho v_t^{*} D^{*}/\mu$, with $v_t^{*}$ the bubble terminal velocity), and the bubble major-to-minor diameter ratio ($\chi$), called aspect ratio. At low $Bo$, the bubble remains nearly spherical and exhibits a steady, rectilinear trajectory~\citep{CGW78,BM95}. As $Bo$ and $Ga$ increase, the bubble becomes more deformed and the wake undergoes a transition from steady to unsteady, leading to path instabilities such as zigzagging or spiraling~\citep{MM02,YPT03,ZM08,CMMT16}. These instabilities are associated with periodic shedding of vortical structures, which vary in symmetry and intensity with bubble shape and velocity, and alter the terminal velocity and lead to abrupt transitions in the bubble’s dynamical regime~\citep{TSG15,CMMT16}.\\

However, most practical situations involve bubbles rising near solid boundaries, such as walls or immersed objects, which significantly affect their motion. This proximity breaks the axial symmetry of the wake, altering pressure and viscous forces and resulting in lateral forces that may either attract or repel the bubble from the wall~\citep{M03c,ST10,ZDCY20}. The interplay between the bubble and a nearby vertical wall is controlled by two opposing forces depending on the value of the control parameters ~\citep{SZM24,Shi24}. The first is an attraction due to the Bernoulli effect, where faster liquid flow in the space between the bubble and the wall pulls them together due to the existence of a pressure minimum in the gap. The second is a repulsion caused by the bubble's wake interacting with the wall, due to vorticity asymmetries across the bubble caused by the presence of the wall. \citet{TTMM02} and \citet{TM03} found experimentally that for $Re \lesssim 35$, the repulsive effect dominates, leading to net migration away from the wall. Above this threshold, the attractive irrotational component becomes significant, often leading to a bouncing behaviour where the bubble approaches and recoils from the wall. This balance between attractive and repulsive forces has also been explored numerically by \citet{ST15}, who confirmed the presence of both effects and emphasised the importance of accounting for bubble deformation and wall-induced wake asymmetries.\\

Recently, \citet{SZM24,SZM25} conducted numerical studies focused on the effect of wall proximity and confinement on bubbles undergoing wake-induced instabilities. Their results confirm that the presence of a wall promotes vortex shedding and the transition to unsteady regimes, and highlight that the magnitude and spatial distribution of the repulsive force strongly depend on the confinement geometry and the relative bubble size. Moreover, they identified by numerical simulations several regimes of bubble motion in the ($Ga, Bo$) plane for a bubble-to-wall distance of one bubble diameter. In the absence of path instability, when $Ga$ and $Bo$ remain below critical thresholds, bubbles may undergo periodic bouncing near the wall or escape the near-wall region after rebounding, driven by a Magnus force arising from rotational flow. For larger $Ga$ and $Bo$,  when path instability is present, certain cases demonstrate a gradual migration away from the wall, whereas others display sustained trapping near the wall accompanied by zigzag motion. \\

Nevertheless, some of the latter scenarios obtained numerically have not been observed experimentally yet. In particular, the BTE (Bouncing-Tumbling-Escaping) regime has been reported experimentally just by \citet{VBW02}. Experimental investigations of bubble-wall interactions remain relatively scarce. Most available studies focus on nearly spherical bubbles at low $Re$, where the migration and wall repulsion effects have been attributed to asymmetries in vortex distribution and irrotational flow interactions~\citep{TTMM02,JP15,JH17}. A few recent works have begun to explore more complex dynamics, but the wall often extends from the gas injector upward, meaning that the bubble is influenced by the wall throughout its entire ascent~\citep{SZM24,ECMBBJ24}. Although there are some experimental works in which the wall edge is placed downwards from the gas injector~\citep{VBW02,TTMM02,TM03}, a systematic study on the effect of the wall distance remains to be done. \\

The present study investigates experimentally a deformable bubble that initially rises freely in an unbounded liquid until it encounters a vertical wall partway along its trajectory. This setup allows for the isolated examination of how the wall modifies the bubble dynamics upon interaction, rather than throughout the entire rising process. The aim is to provide new experimental insights into how the wall presence affects the bubble's rising dynamics when it encounters the wall, as well as to compare the results with the recent numerical studies. This configuration is relevant to many real-world processes where bubbles do not originate next to boundaries but encounter them during their ascent, such as in aeration tanks, multiphase reactors or green hydrogen production. This work aims to complement existing theoretical and numerical studies, bridging the gap between controlled experiments and more complex, practical systems. Moreover, our experimental results can provide new data that may validate and inform future modelling efforts.\\

The paper is organised as follows. Section \ref{sec2} presents the experimental setup, methods and procedure followed to obtain the results. In particular, the observed bubble rising regimes depending on the control parameters are reported and described in Section \ref{sec3}. Finally, Section \ref{sec4} is devoted to the main conclusions of the work. 

\section{Experimental method} 
\label{sec2}

This section presents the experimental methodology, with the facility detailed in Subsection~\ref{sec2.1} and the procedure in Subsection~\ref{sec2.2}.

\subsection{Experimental setup}
\label{sec2.1}

\textbf{\begin{figure*}[tbp]
\begin{center}
\resizebox{1\textwidth}{!}{\includegraphics{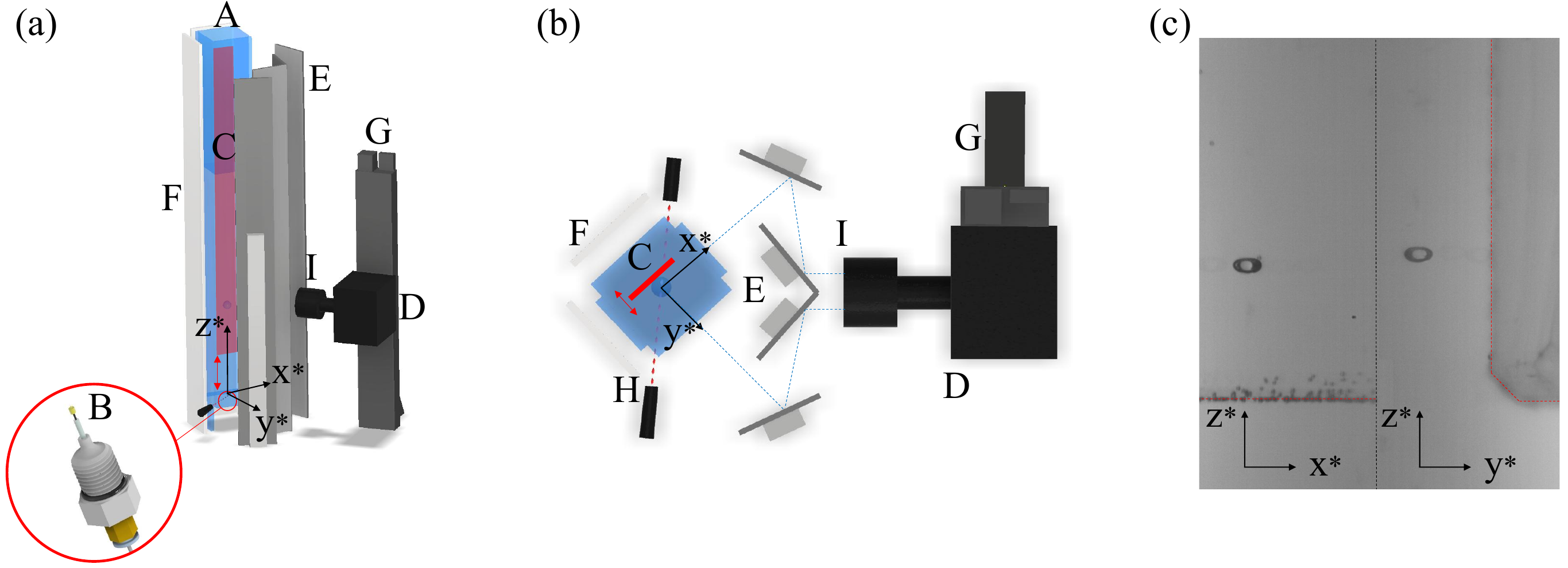}}   
\end{center}
\caption{Experimental setup: three-dimensional scheme (a), top view (b), real image of the experiment (c). Components: tank (A), needle (B), mobile glass wall (C), high-speed camera (D), sets of mirrors (E), LED panel lights (F), vertical motorised rail (G), photo-diode sensor (H), and optical lenses (I). The image on the right side shows the two perpendicular views of the bubbles at the moment when the wall appears in the two views.}
\label{setup}
\end{figure*}}

The experimental setup is sketched in Figure \ref{setup}. We used a parallelepipedal glass tank/column (A in Fig.~\ref{setup}a) to ensure optical access and maintain cleanliness, measuring 10 cm in both length and width and 100 cm in height. We used ultrapure water in all our experiments, provided by a water purification machine ({\sc Direct-Q®3}). The density and viscosity were $\rho=998$ kg/m$^3$ and $\mu=1$ $m$Pa$\cdot$s at $T=293\pm1$ K, with a conductivity of around 3.5 $\mu$S/cm. The gas used in the experiments was nitrogen (=99.998\%, $\rho_g=1.25$ kg/m$^3$, $\mu_g=17.6$ $\mu$Pa$\cdot$s, {\sc Carburos Metálicos SL}). The surface tension of the gas-liquid interface was $\sigma=72$ mN/m, measured using the Theoretical Interface Fitting Analysis ({\sc TIFA})~\citep{FMC07}. Bubbles were generated by injecting a given flow rate of nitrogen through a needle placed at the centre of the tank base (B in Fig.~\ref{setup}a). Note that the reference frame origin is set at the centre of the needle tip (see Fig.~\ref{setup}). Directions $y^*$ and $x^*$ define the horizontal components of the bubble trajectory in the direction perpendicular and parallel to the wall, respectively, while $z^*$ defines the vertical component. To control the bubble size, we used needles with tip inner diameters ranging from 10 to 90 $\mu$m. The needle tips were fabricated using Nanoscribe Photonic Professional GT2 with the dip-in laser lithography (DiLL) configuration. The 25$\times$ objective was dipped into the IP-S resin droplet deposited on an ITO-coated glass substrate. We chose the shell and scaffold writing strategy. The typical slicing and hatching distances were $1$ $\mu$m and 0.5 $\mu$m. The part was developed in $\sim$25 ml of propylene glycol monomethyl ether acetate (PGMEA) for 24 h and then washed in isopropanol for 2 h. Then, the unexposed resin inside the shell was cured for 60 min inside the UV Curing Chamber ({\sc XYZprinting}). \\

\begin{table}
  \begin{center}
\def~{\hphantom{0}}
  \begin{tabular}{lcccccc}
      Exp. case  & $D^*$ (mm) & $Bo$ & $Ga$ & $Re$ & $L$\\[3pt]
\#1 & $0.800\pm0.021$ & $0.09\pm0.005$ & $71\pm3$ & $138\pm8$ & $0.30\lesssim \text{L}\lesssim 5.10$\\
\#2 & $1.146\pm0.001$ & $0.18\pm0.001$ & $122\pm0$ & $340\pm3$ & $0.44\lesssim \text{L}\lesssim 4.36$\\
\#3 & $1.288\pm0.008$ & $0.23\pm0.003$ & $145\pm1$ & $415\pm12$ & $0.39\lesssim \text{L}\lesssim 3.92$\\
\#4 & $1.472\pm0.011$ & $0.29\pm0.004$ & $177\pm2$ & $514\pm6$ & $0.34\lesssim \text{L}\lesssim 3.37$\\
\#5 & $1.565\pm0.001$ & $0.33\pm0.004$ & $194\pm2$ & $556\pm3$ & $0.32\lesssim \text{L}\lesssim 3.21$\\
  \end{tabular}
  \caption{Experiments overview with wall. $D^*$, $Bo$, $Ga$, and $Re$ are mean values from the different distances $L$ assessed.}
  \label{tab:exp}
  \end{center}
\end{table}

A rigid and hydrophilic glass wall ($80\times1000\times6$ mm, with a typical root-mean-square roughness of a few nanometres) was placed vertically inside the tank (C in Figs.~\ref{setup}a,b). The wall was ensured parallel to the $x^*z^*$ plane by using a goniometer with micrometric precision. The vertical and horizontal distances from the wall to the centre of the injector, that is, to the bubble centroid just after its generation, $z_w^{*}$ and $y_w^{*}$, respectively, were controlled with a biaxial translation stage with micrometric precision. In all our experiments developed to study the effect of the wall on the dynamics of the rising bubble, $z_w^{*}$ was fixed at 144 mm to ensure that the bubble had already reached its terminal velocity ($v_t^{*}$) and its stable shape ($\chi$) before it encountered the leading edge of the wall. The horizontal distance $y_w^{*}$ was systematically varied for each case, ranging from 0.25 to 5 mm, to study its effect on the bubble's rising regime.\\

To perform the measurements, a virtual binocular stereo vision system~\citep{LWZGZXLL22, RVCMLH24} was employed to capture two perpendicular views of the rising bubble. This system consisted of a high-speed CMOS camera ({\sc Photron, Fastcam Mini UX50}) that captured two perpendicular views of the bubble (D in Figs.~\ref{setup}a,b). Two sets of mirrors were aligned as illustrated in Fig. \ref{setup} (E in Figs.~\ref{setup}a,b) to visualise both views of the bubble in the centre of their side of the image, placed in a specific position controlled by goniometers. Two LED panel lights (F in Figs.~\ref{setup}a,b) and the camera were positioned on opposite sides of the tank, with a diffuser plate placed between the LED panel light and the tank to ensure uniform illumination. The camera was mounted on a vertical motorised rail (G in Figs.~\ref{setup}a,b) and moved at a constant speed to track the bubble during its ascent. A laser beam and a photo-diode sensor (H in Fig.~\ref{setup}b) detected the bubble pinch-off and triggered the vertical motorised rail. The bubble movement was recorded at 500 fps with an exposure time of 625 $\mu$s, producing images with a resolution of 1024×1280 pixels. The camera was equipped with a set of optical lenses (I in Fig.~\ref{setup}b), including a Nikkor AF 35-70 mm 1:3.3-4.5 zoom objective ({\sc Nikon}), a 2$\times$ zoom, and 20 mm extension rings, achieving a magnification of 21.62 $\mu$m/pixel. All components were mounted on an optical table equipped with a pneumatic anti-vibration isolation system ({\sc Thorlabs}) to eliminate external vibrations from the building. All the experiments were performed at room temperature (20 $^{\circ}$C). \\

The variables involved in the free rise of a bubble are the equivalent bubble diameter, $D^{*}$, the liquid physical properties, $\rho$, $\mu$, $\sigma$, i.e., density, dynamic viscosity, and surface tension, respectively, gravity acceleration, $g$, and the horizontal gap between the wall and the centre of the gas injector, $y^*_w$. Since the bubbles in this study are deformable, the bubble size was characterised using the equivalent diameter $D^{*}$, defined as the diameter of a sphere with a volume, $V^{*}$, identical to that of the bubble, $D^{*} = (6V^{*} / \pi )^{1/3}$. With the needle tips used, the bubble equivalent diameter ranged from $D^{*}=0.8$ to $D^{*}=1.6$ mm, smaller than the critical value, $D^{*}=1.82$ mm~\citep{D95}, therefore, with a stable path (without the effect of a wall). Table~\ref{tab:exp} shows the details of all the experiments performed in this work. Note that, in this manuscript, asterisks denote dimensional parameters, in contrast to their dimensionless counterparts. We take $D^{*}$ and gravitational velocity $(gD^{*})^{1/2}$ as the characteristic length and velocity, respectively to obtain the corresponding dimensionless parameters for our analysis. Then, our study is controlled by three dimensionless parameters: the Galilei number $Ga=\rho (g D^{*3})^{1/2}/\mu$, the Bond number $Bo=\rho g D^{*2}/\sigma$, and the initial dimensionless distance between the bubble centroid and the wall $L$, $L = y_w^{*}/D^{*}$. In our experiments, the Morton number, a dimensionless parameter that helps characterize the shape and behaviour of bubbles moving in a surrounding fluid, has a constant value ($Mo = g\mu^{4}/(\rho \sigma)^{3}=Bo^{3}/Ga^{4}=2.64\times 10^{-11}$ because it depends only on the properties of the liquid. Then, $Bo$ is related with $Ga$ as $Bo=Mo^{1/3} Ga^{4/3}=1.38\times 10^{-11/3} Ga^{4/3}$. Moreover, as a result of the experiments, we obtained the bubble terminal velocity, $v_t^*$, which allowed us to calculate the Reynolds number, $Re=\rho v^*_t D^*/\mu$. The values of the Bond, Galilei and Reynolds numbers lie in the intervals $0.09\lesssim \text{Bo}\lesssim 0.33$, $71\lesssim \text{Ga}\lesssim 194$ and $138\lesssim \text{Re}\lesssim 556$ (see Table \ref{tab:exp}). The dimensionless horizontal distance from the wall to the injector was varied from $0.3 \lesssim L \lesssim 5$.

\subsection{Experimental procedure}
\label{sec2.2}

To inject the gas, we used a small-volume 100 $\mu$L syringe ({\sc Hamilton 700 series}) driven by a syringe pump ({\sc Harvard Apparatus PHD 4400}) to guarantee constant flow rate conditions, and very small nitrogen flow rates were supplied to ensure quasi-static bubble formation. In fact, high-speed video images of the bubble formation were analysed to check this condition~\citep{RVCMLH24}. Moreover, before each experiment, we placed the corresponding tip for the size of the bubble to be studied at the base of the tank, and the tank was thoroughly cleaned by filling and draining it repeatedly, with ultrapure water. Immediately afterwards, it was filled with the working liquid. An experimental run was then conducted by first positioning the wall vertically at the higher position of $y_w^{*}$ and bringing it closer to the injector (initial position of the bubble) by using the biaxial translation stage and a goniometer with micrometric precision. We repeated this procedure for each tip, that is, each bubble size. \\

%In each experiment, nitrogen was injected through the needle to form the bubble in the centre of the tank bottom. The bubble rose across the tank until it reached the free surface. 

\begin{figure}[tbp]
\begin{center}
\resizebox{0.4\textwidth}{!}{\includegraphics{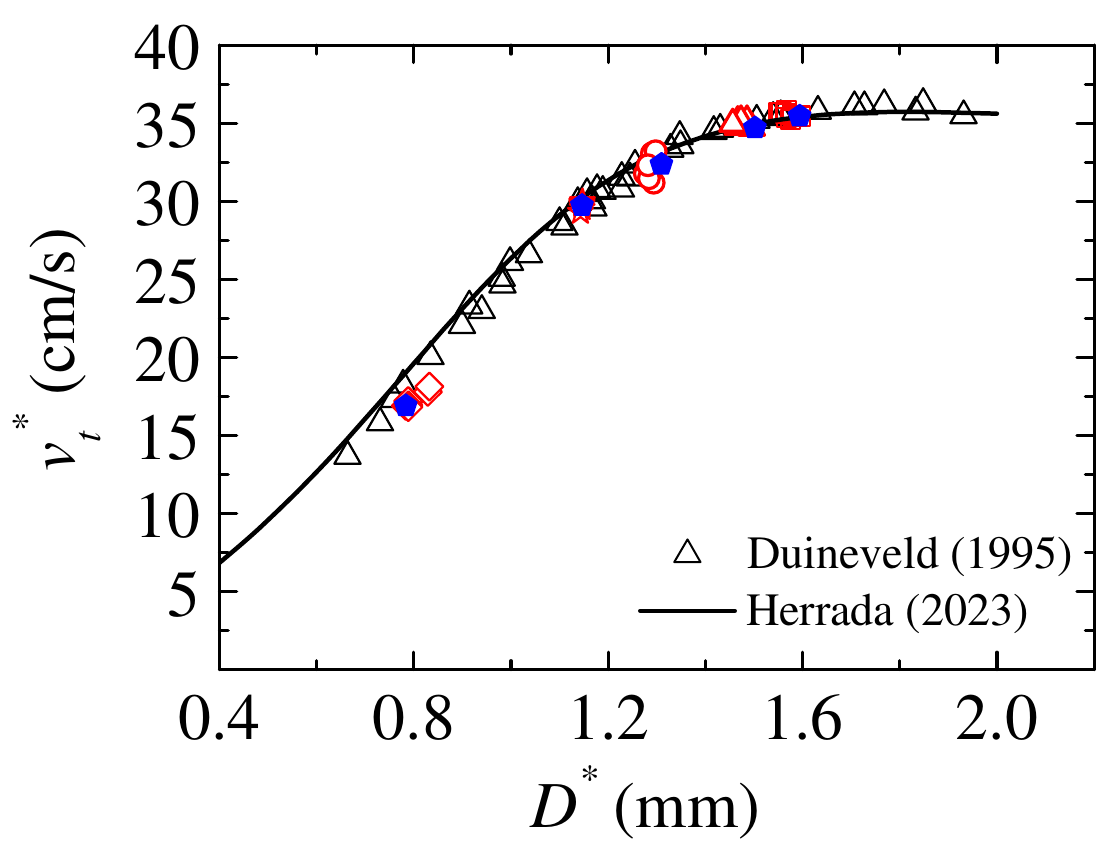}}   
\end{center}
\caption{Bubble terminal velocity $v_t^{*}$ as a function of the diameter $D^{*}$ for clean water. Black symbols and line correspond to experimental results of~\citet{D95} and the numerical predictions of~\citet{HE23} without wall, respectively. Blue symbols stay for our results without vertical wall, and red symbols depict our results with wall at $z_w^{*}=144$ mm and $y_w^{*}$ ranging from 0.3 to 5 mm.}
\label{vt_R}
\end{figure}

The images were processed at the pixel level using an in-house {\sc Matlab code} to determine the bubble centroid positions, as well as its shape and velocity. Additional information is available in the supplemental materials of \citet{RVCMLH24}. Each experiment was repeated five times. The results are the values averaged over the five experimental realisations, being the experimental uncertainty quantified by the standard deviation. The bubble adopts essentially the same path in all the cases. The small inevitable differences between the initial conditions of the five trajectories hardly affect the shape of the path. \\

We performed preliminary experiments without a wall to validate our experimental procedure. We measured the terminal velocity $v^*_t$ for different bubble sizes, calculated as the time-average vertical velocity, $v^*_z,$ once it reaches a stable value. The results (see Fig. \ref{vt_R}, blue symbols) agree remarkably with the experimental results of \citet{D95} and the numerical predictions of \citet{HE23}. This agreement confirms that our experiments correspond to clean water.

\section{Results and discussion} \label{sec3}

In this section, we first describe the different regimes observed as a function of the wall separation for various values of $Bo$ (or equivalently $Ga$) in subsection~\ref{sec3.1}. Subsequently, subsection~\ref{sec3.2} presents a regime map in the dimensional parameter space $(Bo, L)$ and $(Ga, L)$, which enables identification of the transitions between the different behaviours.

\subsection{Description of the rising regimes} \label{sec3.1}

\begin{figure*}[tbp]
\begin{center}
%figa-tip75-nowall-D1.48 infty
%figb-tip75-3mm-D1.48L2.05
%figc-tip90-3mm-D1.58L1.93
%figd-tip75-1mm-D1.48L0.68
%fige-tip10-1mm-D0.79L1.27
%figf-tip10-3mm-D0.79L3.8
\resizebox{0.32\textwidth}{!}{\includegraphics{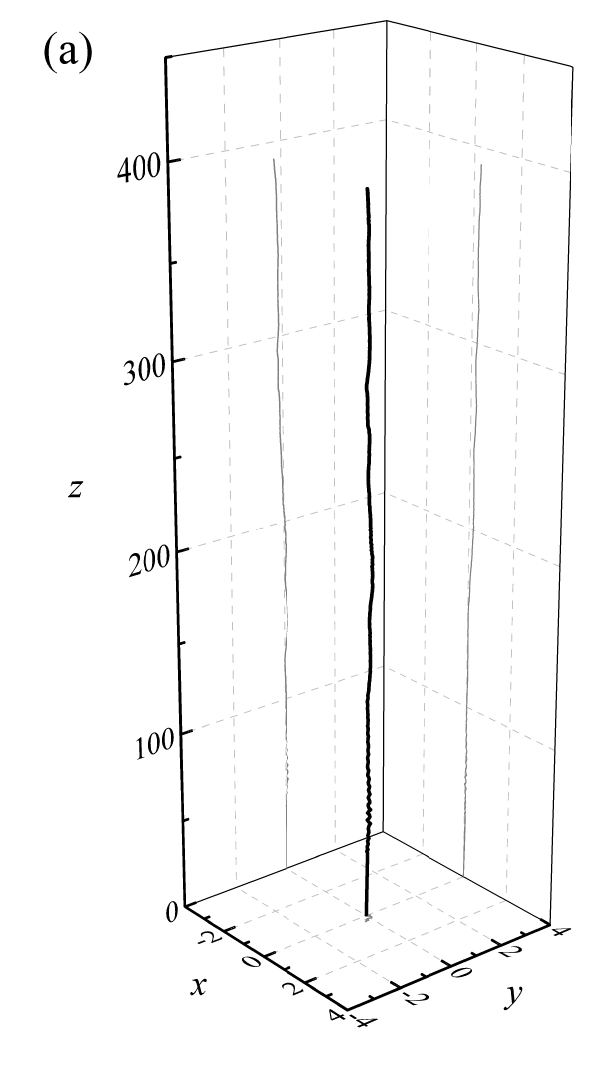}}
\resizebox{0.32\textwidth}{!}{\includegraphics{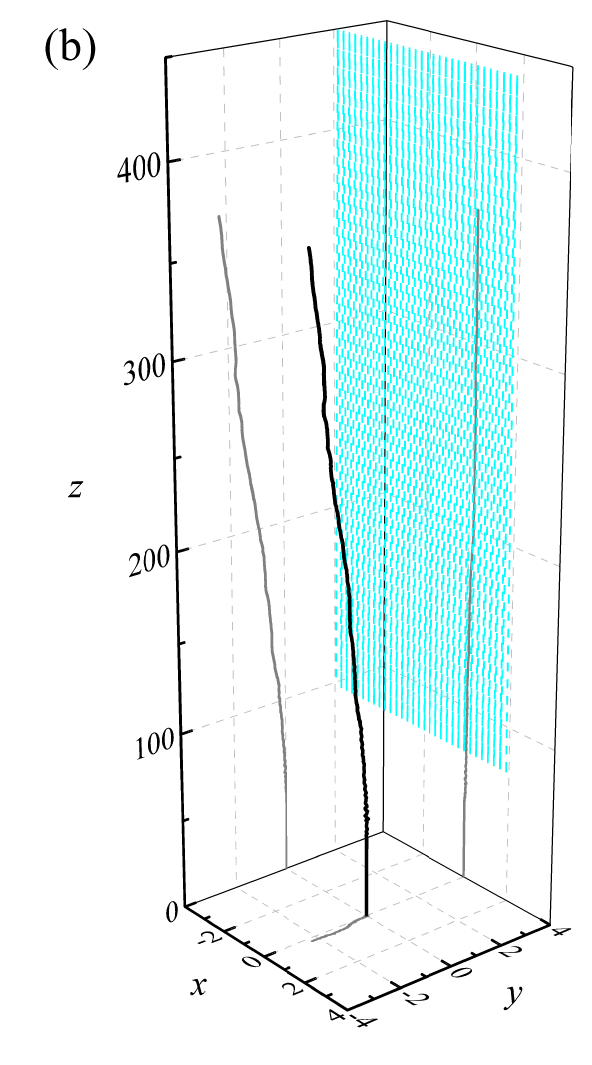}}
\resizebox{0.32\textwidth}{!}{\includegraphics{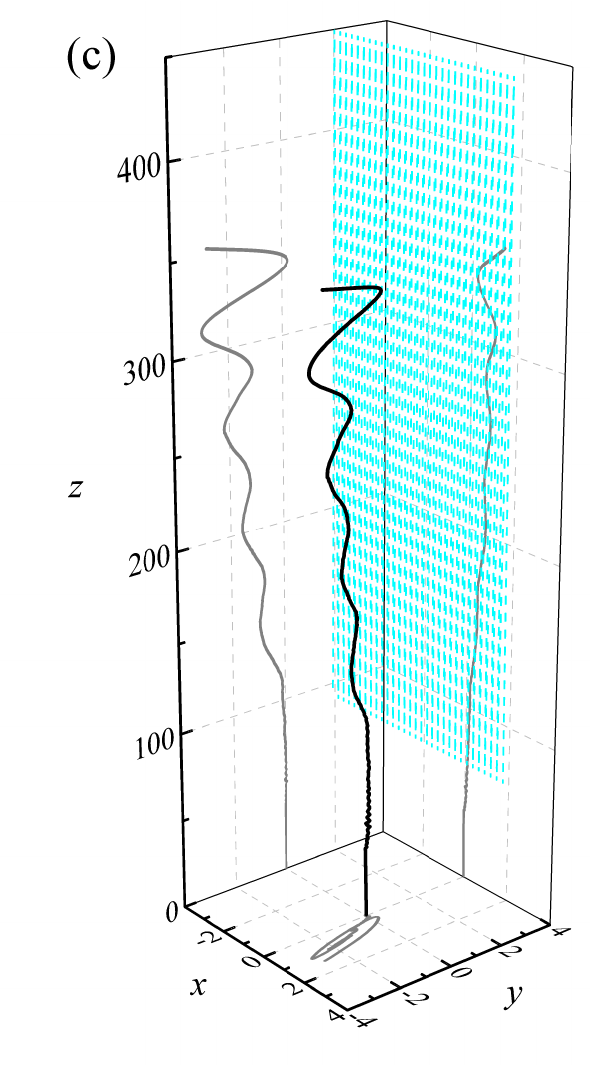}}
\resizebox{0.32\textwidth}{!}{\includegraphics{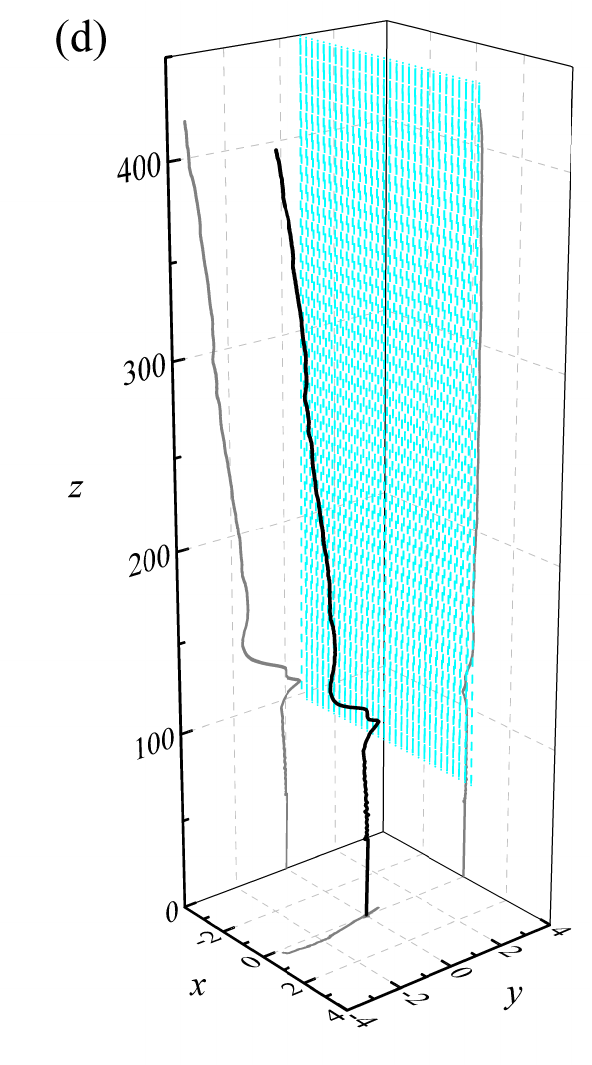}}
\resizebox{0.32\textwidth}{!}{\includegraphics{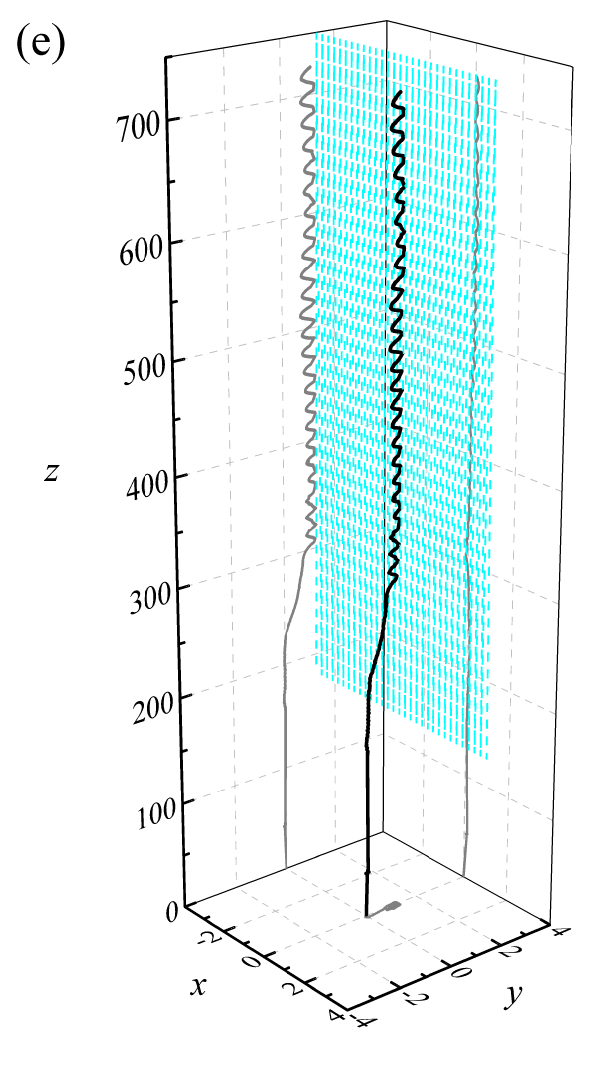}}
\resizebox{0.32\textwidth}{!}{\includegraphics{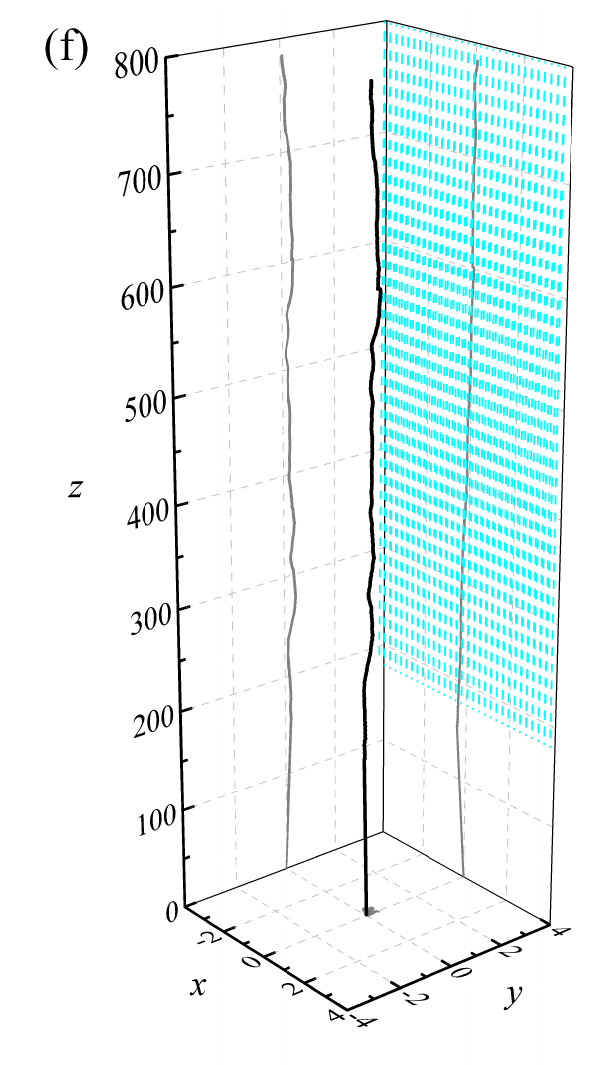}}
\end{center}
\caption{Three-dimensional bubble path, together with its projections onto the planes $(x,y)$, $(x,z)$, and $(y,z)$, showing the different rising regimes observed in this work: (a) Rectilinear path (RP): Exp. \#4, $L=\infty$ (no wall); (b) Migration Away (MA): Exp. \#4, $L=2.02$; (c) Migration Away (MA) with path instability: Exp. \#5, $L=1.93$; (d) Collision and migration away (C+MA): Exp. \#4, $L=0.68$; (e) Periodic collisions (PC): Exp. \#1, $L=1.27$; (e) Rectilinear path (RP): Exp. \#1, $L=3.8$. The blue area represents the vertical wall.}
\label{trajectories}
\end{figure*}

Figure \ref{trajectories} illustrates the four different types of trajectories or behaviours observed due to the influence of a downstream-located wall, varying the initial wall-bubble horizontal distance $y_w^{*}$ ($L$) and the bubble size $D^{*}$ ($Bo$ and $Ga$). The reference case without wall for $D^{*}=1.48$ mm ($Bo=0.3$, $Ga=181$, $L=\infty$) is also shown in Fig. \ref{trajectories}(a). The latter follows a stable (straight or rectilinear) path, with negligible oscillations likely attributable to camera vibrations, as these fluctuations were absent when the camera remained stationary. Once the vertical wall is present, we observed different behaviours, which have been classified into four regimes as follows:\\

\begin{figure*}[tbp]
\begin{center}
\resizebox{0.9\textwidth}{!}{\includegraphics{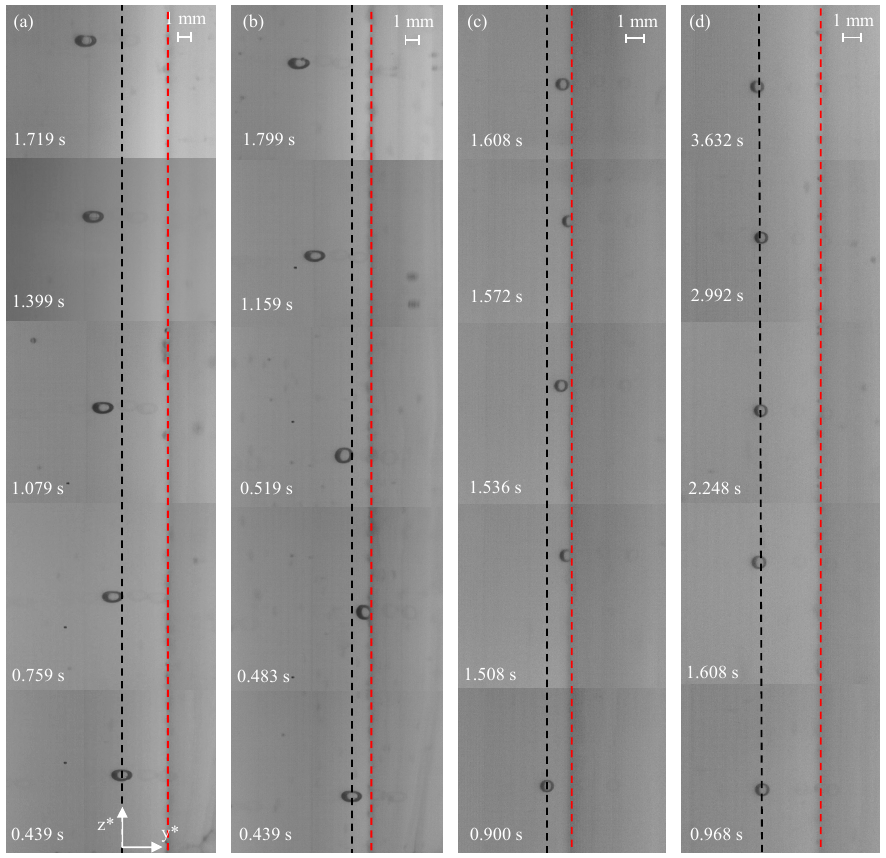}}   
\end{center}
\caption{Images of the bubble rising associated with the different behaviours found, from left to right: migration away (MA), collision and migration away (C+MA), periodic collisions (PC), and rectilinear path (RP). The bubble ejection occurs at $t=0 s$. The white dashed line serves as a guide, indicating a vertical path, and the red dashed line is the glass wall edge (see the corresponding videos in the Supplemental Material).}
\label{images_behaviours}
\end{figure*}

\begin{itemize}

    \item Migration away from the wall (MA): after an initial acceleration phase, the bubble encounters a wall positioned at a height where its terminal velocity has been clearly established. The bubble exhibits a regular repulsive migration in the $yz$ plane. That is, the bubble keeps its stable behaviour, but a repulsive effect is induced by the wall. This regime is one of three distinct types of near-wall bubble motion observed in moderately inertial regimes, and has been reported previously, especially for larger $Bo$~\citep{ECMBBJ24,SZM24,SZM25}. The underlying physical mechanism driving this movement is primarily vortical, originating from the interaction between the bubble’s wake and the vertical wall. As the bubble ascends, it displaces fluid laterally within its wake. When this displaced fluid reaches the wall, the wall induces a slight lateral flow away from itself, creating a pressure gradient that generates a lateral force that pushes the bubble away from the wall. The near wake of the bubble has been shown numerically~\citep{SZM24} to remain nearly axisymmetric, but, due to the interaction between the wake and the wall, a weak streamwise vorticity develops along the wall and at the rear of the bubble. Figures \ref{trajectories}(b) and \ref{images_behaviours}(a) show this first behaviour for Exp. \#4 ($Bo=0.29$ and $Ga=177$, with $L=2.02$). In some cases, for $Bo\gtrsim0.33$, the migration away motion is accompanied by the development of path instability (planar zigzag that evolves to a flattened spiral trajectory~\cite{CTMM16}), see Fig. \ref{trajectories}(c) for Exp. \#5 ($Bo=0.33$ and $Ga=194$, with  $L=1.93$). This destabilising effect of the wall proximity has been reported numerically~\citep{ZDCY20,YZLZZL22,SZM24,SZM25}.\\
    
    \item Collision and migration away (C+MA): if $L$ is decreased from that in MA regime, the bubble initially approaches the wall, undergoes a sudden direct collision against the vertical wall, eventually escapes from the near-wall region, and subsequently experiences a repulsive migration from the wall along its trajectory. Figures \ref{trajectories}(d) and \ref{images_behaviours}(b) illustrate this second behaviour for Exp.\#4 ($Bo=0.3$ and $Ga=176$) with $L=0.68$. Note that, after escaping, the bubble reaches a large final wall-normal position ($y\simeq 4$ at $z\simeq 400$). This regime has been barely explored, being reported experimentally by \citet{VBW02}. Nevertheless, they attributed this behaviour to a slight tilt of the wall. This regime could be related to the Bouncing-Tumbling-Escaping (BTE) regime recently reported by \citet{SZM25}. They showed this behaviour to occur under highly inertial conditions, specifically for $Bo\lesssim 0.2$ at $Ga$ = 70 and $Bo \lesssim 0.4$ for $Ga \gtrsim 198$. These conditions place the BTE regime below the neutral curve for path instability, where the bubble would otherwise rise vertically in an unbounded fluid. Bubbles within this regime exhibit high Reynolds numbers and undergo limited to moderate deformation, with typical aspect ratios $\chi \lesssim 1.85$. \citet{SZM25} demonstrated that the lateral escape of bubbles in the BTE regime is primarily driven by a Magnus-like lift force. This force originates from the intense rotational flow generated around the bubble during its collision with the wall. As the bubble rebounds, part of its translational kinetic energy is converted into a spinning motion of the surrounding fluid, quantified by a non-zero interface-averaged spinning rate. The resulting lift force, acting perpendicular to both the local spin and rise velocity, opposes the wall-induced attraction and becomes strong enough to displace the bubble away from the near-wall region, enabling its eventual escape.\\

    \item Periodic collisions (PC): for smaller bubbles and when the wall is sufficiently close, the bubble approaches the wall and exhibits a sequence of periodic collisions along its vertical ascent. Figures~\ref{trajectories}(e) and \ref{images_behaviours}(c) illustrate this behaviour for Exp. \#1 ($Bo = 0.08$, $Ga = 69$, $L = 1.27$), where the bubble repeatedly bounces off the wall with approximately constant amplitude and frequency. This regime has been observed previously at sufficiently high Galilei numbers and low Bond numbers~\citep{TM03,VBW02,SZM24}. The mechanism underlying these periodic interactions involves a balance between attraction and repulsion near the wall. Initially, the bubble is laterally drawn toward the wall due to the Bernoulli effect, which lowers the pressure in the gap between the bubble and the wall. However, as the bubble approaches and rebounds, its wake interacts with the wall, generating a repulsive force. This force arises primarily from the periodic shedding of streamwise vortices induced by the strong shear near the wall. These vortices produce a lateral pressure gradient that counteracts viscous resistance and drives the bubble away from the wall. The interaction ultimately results in a quasi-harmonic motion about an equilibrium position, where the bubble oscillates with low to moderate deformation~\citep{SZM24}. \\
        
    \item Rectilinear path (RP): when the initial wall separation $L$ is sufficiently large compared to that in the PC regime, the bubble rises along a steady, rectilinear trajectory parallel to the wall, maintaining an approximately constant lateral distance $y$. Under these conditions, the influence of the wall on the bubble motion becomes negligible, and the bubble behaves essentially as it would in an unbounded fluid. This behaviour is illustrated in figures~\ref{trajectories}(f) and \ref{images_behaviours}(d) for Exp. \#1 ($Bo = 0.08$, $Ga = 69$, $L = 3.8$), where the bubble trajectory closely matches that observed in the unbounded case (Fig.~\ref{trajectories}a), indicating an absence of near-wall hydrodynamic interactions.

\end{itemize}

\begin{figure}[tbp]
\begin{center}
\resizebox{0.49\textwidth}{!}{\includegraphics{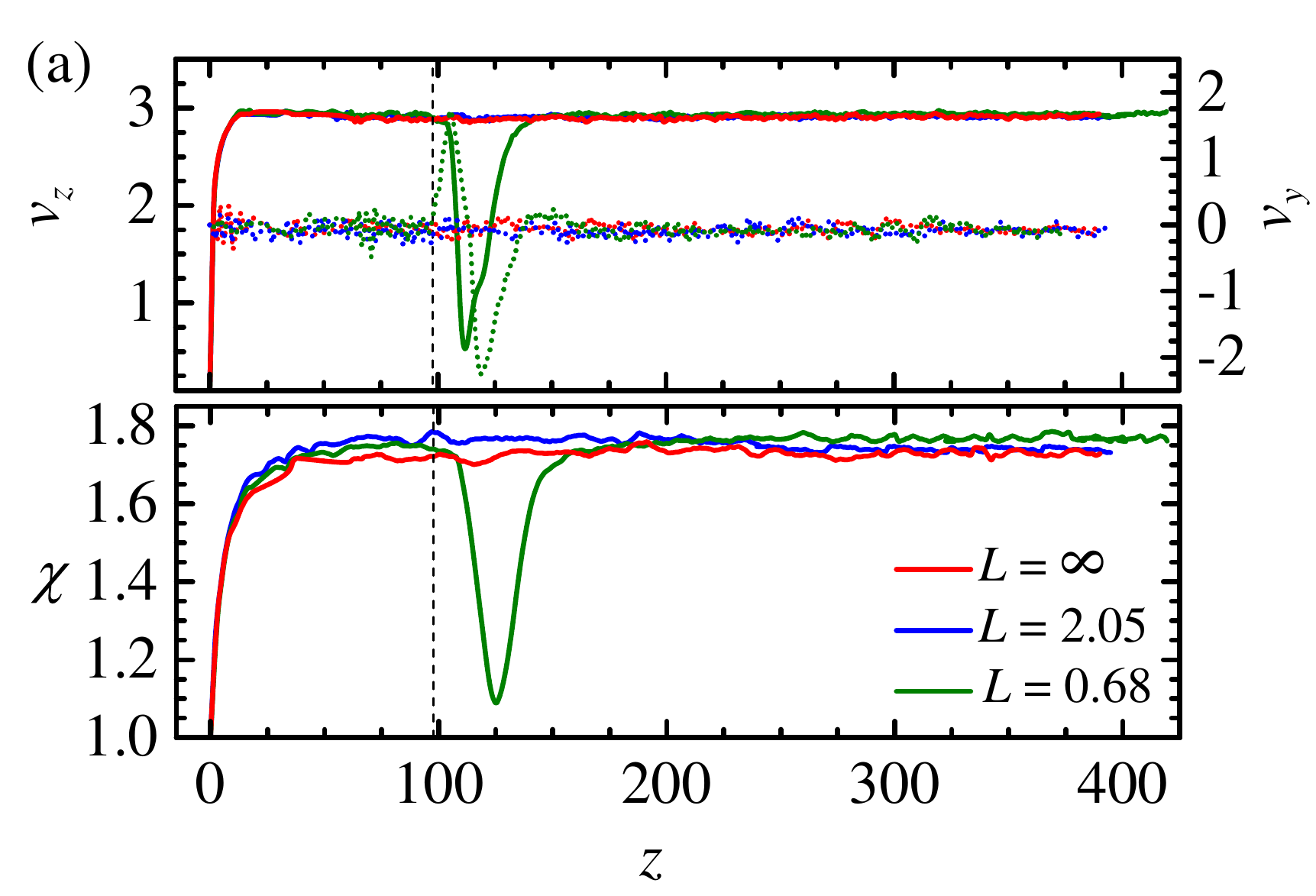}}
\resizebox{0.49\textwidth}{!}{\includegraphics{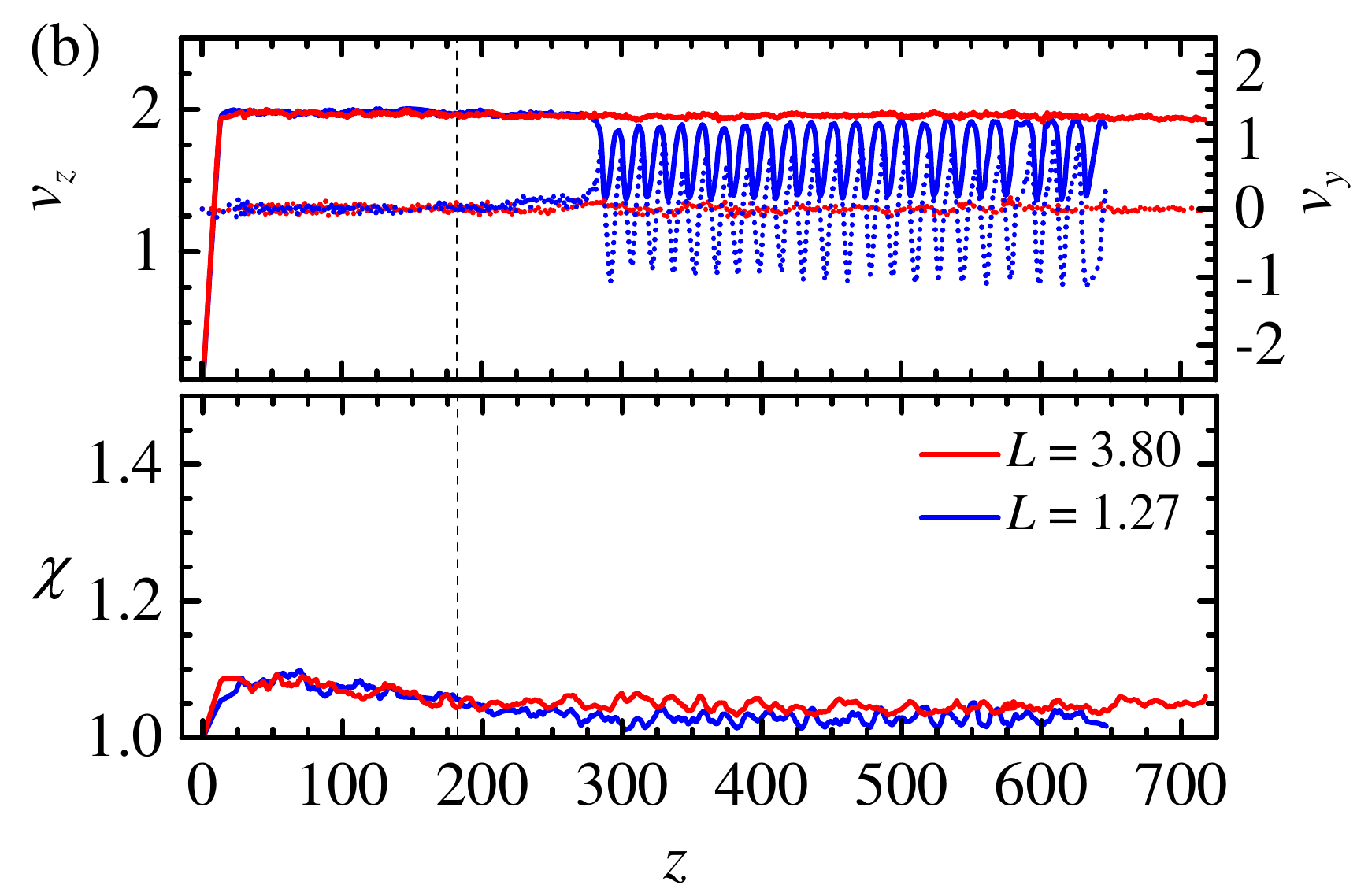}}
     \end{center}
\caption{Dimensionless vertical bubble velocity $v_z$ (left axis), wall-normal bubble velocity $v_y$ (right axis), and aspect ratio $\chi$ as a function of the dimensionless vertical position $z$ of the bubble centroid corresponding to the four regimes. (a) Exp. \#4: unbounded flow (red), MA (blue) and C+MA (green). (b) Exp. \#1:  RP (red) and PC (blue). The wall was positioned at the same position above the tip of the feeding capillary in all cases, $z_w^{*}=144$ mm (vertical dashed line).}
\label{vz_X}
\end{figure}

For further analysis of these four type of trajectories or behaviours, Fig. \ref{vz_X} depicts the dimensionless bubble velocity (vertical, $v_z$, and wall-normal, $v_y$ components), and aspect ratio $\chi$ as a function of the dimensionless vertical position $z$ of the bubble centroid. Note that the wall-normal velocity is much lower than the vertical one for all the cases, that is, the migration effect is low. In particular, for the MA regime (blue line in Fig. \ref{vz_X}a), both the evolution and values of velocities and $\chi$ over $z$ are practically the same as those without wall ($L=\infty$, red line), reaching a stable value around $v_z\simeq3$, $v_y\simeq 0$ and $\chi\simeq1.75$. In most of the cases of MA analysed here, the tilt angle of the path was around $0.5^\circ$, independently of the value of $L$, different from what was observed by \citet{ECMBBJ24} for higher $Bo$ number bubbles, who reported higher angles and a dependency on $L$. \\

Nevertheless, for the C+MA regime (green line in Fig.~\ref{vz_X}a), both the vertical velocity $v_z$ and the bubble aspect ratio $\chi$ exhibit a marked decrease exclusively during the collision event, reaching minimum values of $v_z \simeq 0.5$ and $\chi \simeq 1.1$, respectively. This behaviour reflects the strong deceleration and deformation induced by the wall interaction. Shortly after the collision, both $v_z$ and $\chi$ recover rapidly, returning to their pre-collision values ($v_z \simeq 3$, $\chi \simeq 1.75$), consistent with the dynamics observed in the absence of the wall. The evolution of the wall-normal velocity $v_y$ provides further insight into the sequence of events. As the bubble approaches the vertical position of the wall, $v_y$ increases sharply, indicating lateral acceleration toward the wall. This acceleration phase coincides with the deceleration in the vertical direction, marking the onset of interaction. The collision occurs when $v_y$ abruptly drops to zero, which coincides with the minimum in $v_z$. Immediately thereafter, $v_y$ becomes negative, signifying that the bubble is repelled from the wall. This repulsion intensifies rapidly, with the wall-normal velocity reaching a minimum of approximately $v_y \simeq -2.27$. Notably, the maximum deformation (i.e., the minimum $\chi$) coincides with this moment, consistent with observations in the BTE regime reported by \citet{SZM25}. As the bubble migrates away from the wall, the magnitude of $v_y$ progressively decreases, eventually settling at a small negative value ($v_y \simeq -0.08$). This steady migration phase corresponds to the escaping branch described by \citep{SZM25}, driven by a Magnus-like lift force generated from the rotational flow around the bubble induced during the collision. The overall dynamics—including the abrupt changes in $v_y$, the recovery of $v_z$ and $\chi$, and the emergence of a sustained wall-normal drift—are in excellent qualitative agreement with the BTE regime dynamics described in \citep{SZM25}.\\

In the rectilinear path (RP) regime (red line in Fig.~\ref{vz_X}b), the bubble follows a steady vertical trajectory parallel to the wall, maintaining an approximately constant lateral distance $y_w$. Under such conditions, the influence of the wall becomes negligible, and no significant variations in $v_z$, $v_y$, or $\chi$ are observed. Nevertheless, for the PC regime, (blue line in Fig.~\ref{vz_X}b), the bubbles are smaller and practically spherical ($\chi$ around 1.05). As a result, the successive collisions with the wall do not significantly affect the bubble shape. However, both velocity components exhibit clear periodic oscillations that reflect the repeated bouncing behaviour characteristic of this regime. After reaching the wall position (at $z \simeq 180$), the vertical velocity $v_z$ begins to oscillate, decreasing from approximately the value for unbounded flow ($v_z \simeq 1.9$) to a minimum of $v_z \simeq 1.4$ during each impact. The modulations in $v_z$ are tightly coupled to the evolution of $v_y$, which periodically reverses sign. Specifically, $v_y > 0$ as the bubble approaches the wall, vanishes at the instant of contact ($v_y = 0$), and becomes negative during the rebound, indicating migration away from the wall. The minimum in $v_z$ coincides precisely with the moment of impact, when $v_y = 0$, consistent with the hydrodynamic deceleration reported in \citet{SZM24}. After the collision, $v_y$ becomes increasingly negative, reaching a maximum magnitude of $v_y \simeq -1.24$ during the retreat phase, while $v_z$ recovers. This corresponds to the stage where the bubble moves laterally away from the wall, with the wall-normal velocity decreasing in magnitude as it approaches its maximum displacement. Once the wall-normal velocity vanishes again ($v_y = 0$), the bubble reaches its furthest excursion from the wall. This point also coincides with the local maximum in $v_z$. The subsequent approach phase is characterised by an increase in $v_y > 0$, which peaks midway through the lateral return, before decreasing again as the bubble nears the wall. During this interval, $v_z$ gradually decreases until the next collision occurs. This periodic modulation of both velocity components is in close agreement with the dynamics described in \citet{SZM24}, where the interplay between near-wall shear and wake-induced forces leads to regular vortex shedding and a sustained oscillatory trajectory. \\

\begin{figure} [tbp]
\begin{center}
\resizebox{0.4\textwidth}{!}{\includegraphics{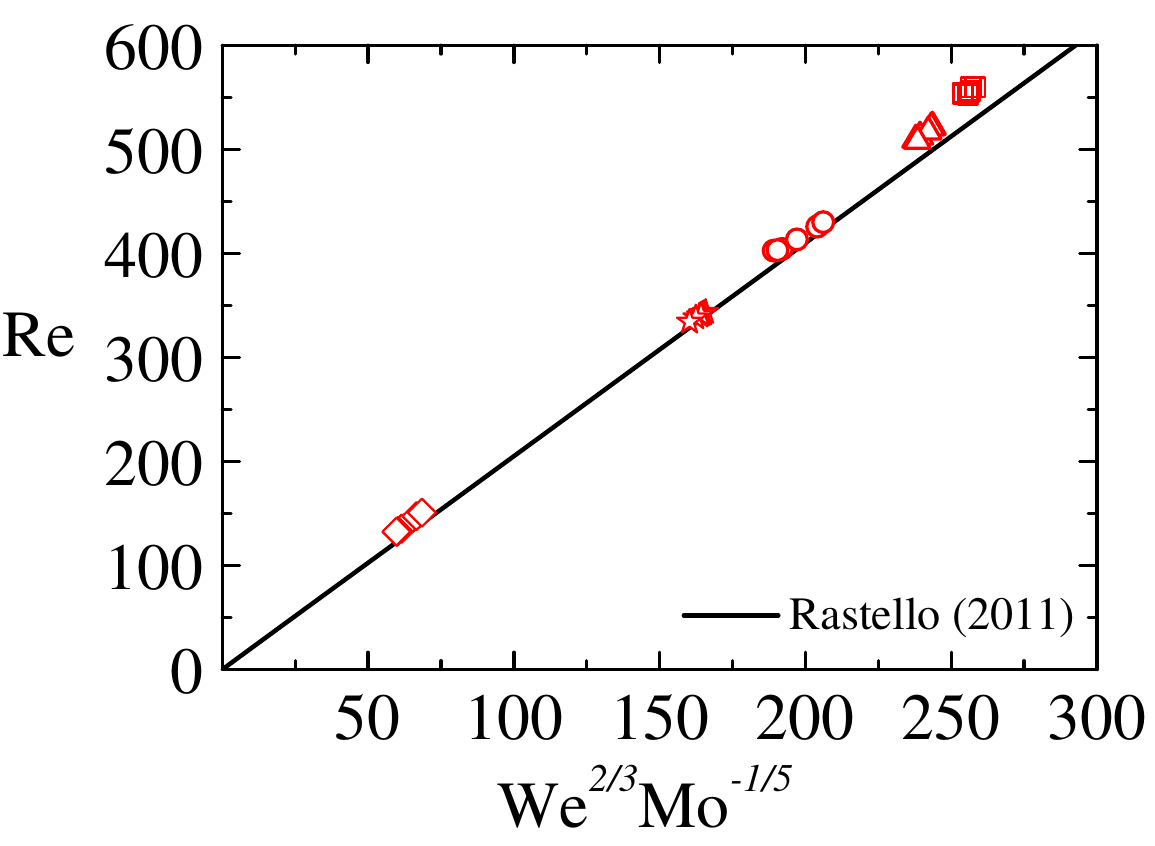}}
\end{center}
\caption{Map of $Re$ versus $We^{2/3}Mo^{-1/5}$. Experimental results for $D^{*}=0.79$, $D^{*}=1.15$, $D^{*}=1.28$, $D^{*}=1.48$ and $D^{*}=1.58$ mm are depicted by diamonds, stars, circles, triangles and squares symbols, respectively. Solid line is a correlation curve from \citet{RML11}, for stable bubbles in clean liquids.}
\label{Re_WeMo}
\end{figure} 

As a conclusion of the present analysis, it may be stated that the bubble terminal velocity, $v_t^*$, is not significantly influenced by the presence of the vertical wall within the range of bubble sizes examined in this study. Figure~\ref{vt_R} (red symbols) illustrates this result. In the case of the periodic collision (PC) regime, the estimation of $v_t^*$ was based on the average of the local maxima of the vertical velocity $v_z$ following each collision. This choice is motivated by the fact that, according to both our observations and previous studies~\citep{SZM24}, the vertical velocity decreases sharply only during the collision phase and subsequently recovers rapidly. Therefore, the post-collision peak of $v_z$ represents a reliable approximation of the effective rise speed the bubble would attain in the absence of repeated wall interactions. From a dimensionless point of view, we have also compared our results with a correlation curve from \citet{RML11}, $Re=2.05 We^{2/3}Mo^{-1/5}$, for stable bubbles in clean liquids, see Fig. \ref{Re_WeMo}. Note that there is a remarkable agreement with this correlation, thus, $Re$ and $We$, $We=\rho (v^*_t)^2 D^*/\sigma =Bo(Re/Ga)^2$, do not change with respect to their homologous cases without the effect of a vertical wall. \\

\begin{figure*} [tbp]
\begin{center}
\resizebox{0.49\textwidth}{!}{\includegraphics{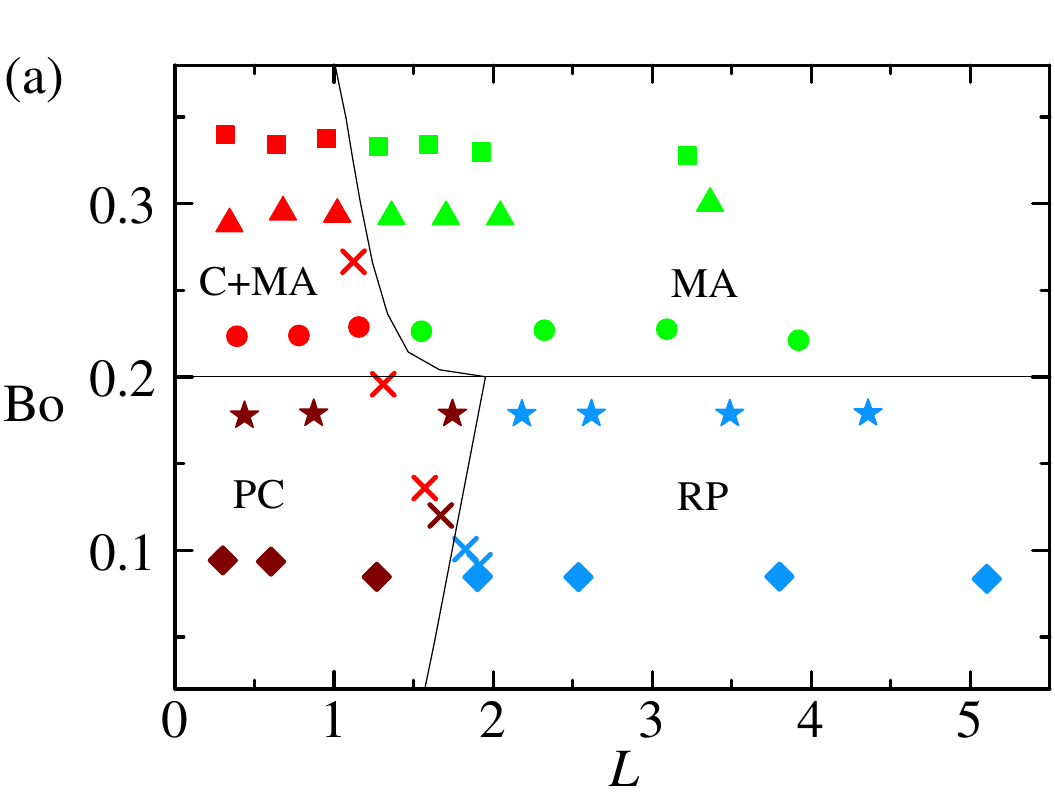}}
\resizebox{0.49\textwidth}{!}{\includegraphics{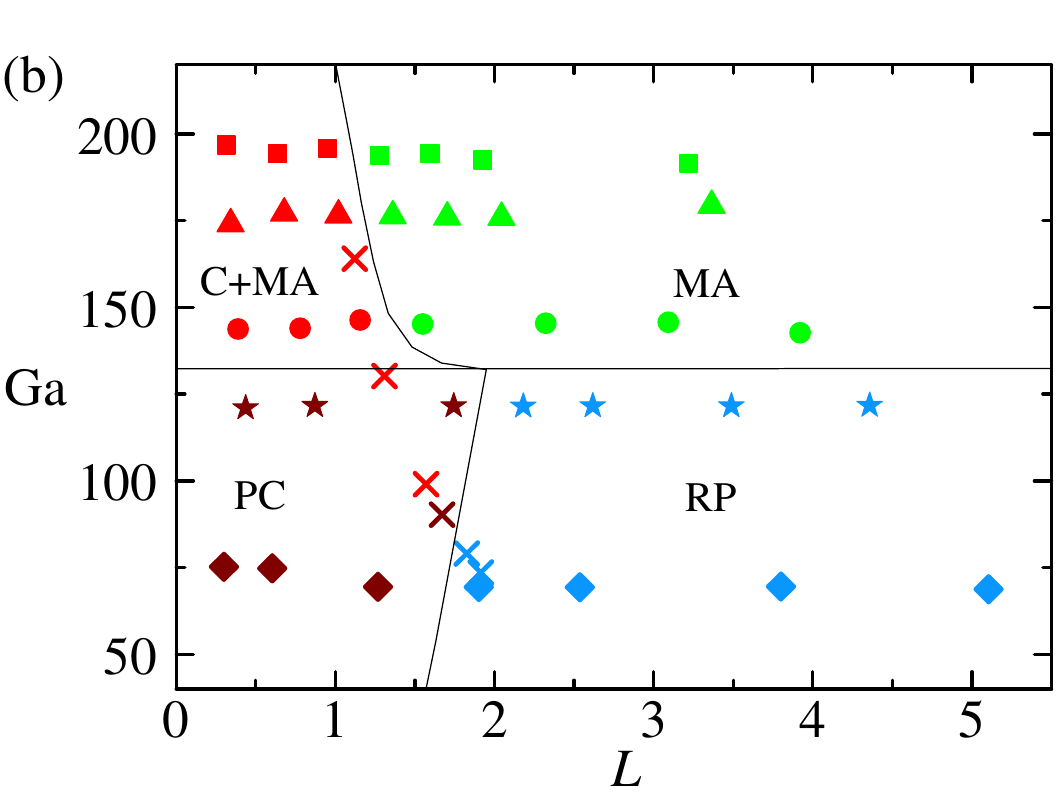}}
\end{center}
\caption{Map of (a) Bond and (b) Galilei number with respect to the initial dimensionless distance between the bubble centroid and the wall $L$. Experimental results for $D^{*}=0.79$, $D^{*}=1.15$, $D^{*}=1.28$, $D^{*}=1.48$ and $D^{*}=1.58$ mm are depicted by diamonds, stars, circles, triangles and squares symbols, respectively. Green symbols correspond to migration away from the wall (MA) region,  red points correspond to collision and migration away (C+MA) region, brown points correspond to periodic collisions (PC) region, and blue points correspond to rectilinear path (RP) region. Blade symbols are results from \citet{VBW02}.}
\label{map}
\end{figure*}  

\subsection{Regimes map and transitions} \label{sec3.2}

The systematic variation of $L$ has allowed for the precise mapping of regime transitions in the \((Bo, L)\) and \((Ga, L)\) parameter spaces, providing insight into the underlying fluid dynamic behaviour. Figure \ref{map} shows a region map of Galilei and Bond numbers with respect to the initial dimensionless distance between the gas injector centre and the wall. To the best of our knowledge, it is the first time that this type of map has been obtained experimentally, sweeping a relatively wide range of $L$. Four regions or regimes are clearly distinguished: green symbols correspond to migration away from the wall (MA) region, red markers correspond to collision and migration away (C+MA) region, brown symbols correspond to periodic collisions (PC) region, and blue points correspond to rectilinear path (RP) region. Blade symbols are results from \citet{VBW02}, which are totally in line with our findings. Note that in the latter work, a systematic variation of $L$ was not explored, in fact, the dimensional distance $y_w^*$ was fixed.\\

As we mentioned previously, there is a lateral force balance owing to two opposite mechanisms: the attractive Bernoulli mechanism resulting from the acceleration of the flow in the gap with the wall, and the repulsive vortical mechanism due to the wall-induced asymmetry of the flow field. For $Bo\lesssim0.2$ ($Ga\lesssim130$), the bubbles are almost spherical ($\chi\simeq$1.2) and the vortical mechanism is weak. If $L$ is high enough, the Bernoulli effect can not activate, and the bubble rises with a rectilinear path (RP), just as if it were an unbounded flow. Nevertheless, if the $L$ is low enough, a transition to the PC regime takes place. In this regime, the wall proximity induces a balance between forces that attract the bubble towards the wall and those that repel it. Bubbles are initially attracted to the wall by the Bernoulli mechanism, but then, a repulsive force, linked to the bubble's wake interaction with the wall, overcomes viscous resistance, and drives the bubble away from the wall. The value of $L$ for this transition increases as $Bo$ (or $Ga$), that is, the PC mechanism requires a higher $L$ to activate as $Bo$ increases because the wake-wall interaction becomes stronger. For this regime (PC), the Strouhal number ($St=fD^*/v^*_t$)) associated with the oscillation frequency ($f$) seems independent of the dimensionless parameter $L$. Taking into account the results showed by \citet{VBW02}, a decreasing trend of $St$ (oscillation frequency) with respect to $Bo$ (bubble size) is observed, reaching values from $St=0.018$ ($f=8.72$ Hz) to $St=0.0004$ ($f=0.93$ Hz), for $Bo=0.09$ and $Bo=0.18$, receptively. However, a more detailed investigation of this regime would be required to confirm this trend, but this issue is beyond the scope of this experimental work. \\

For $Bo\gtrsim0.2$ ($Ga\gtrsim130$), a stronger vorticity at the bubble surface (hence, the wake) produces a larger repulsive force, which gives rise to a migration away regime (MA) when $L$ is large enough. However, when the wall-distance is low enough, a transition to the C+MA regime takes place. Specifically, the transition from MA to C+MA depends on $L$, following approximately a potential law: the value of $L$ at the transition increases when $Bo$ ($Ga$) decreases. Below that critical $L$, Bernoulli mechanism overcomes the vortical mechanism, and a single collision takes place, followed by a migration away from the wall (C+MA), because the distance $L$ just after the collision is higher than the critical one, being the Magnus effect generated by the flow rotation after the collision able to overcome the attraction forces, enabling the bubble to escape from the near-wall region~\citep{SZM25}. \\

For a given $L$, when it is large enough, the Bernoulli mechanism is not active. Thus, we have a transition from migration away (MA) when $Bo\lesssim 0.2$ ($Ga\lesssim130$), in which the wall-bubble wake interaction dominates, to a rectilinear path (RP) for lower $Bo$, in which the wake is too weak, because the bubbles are almost spherical ($\chi$ around 1.2). $Bo\simeq0.2$ ($Ga\simeq130$) also marks the transition from C+MA to PC for a given $L$, when it is low enough. For $Bo\gtrsim0.2$ ($Ga\gtrsim130$), the repulsive mechanism resulting from the interaction of the bubble wake with the wall beats the Bernoulli mechanism. However, when $Bo\lesssim0.2$ ($Ga\lesssim130$), a transition to the PC regime takes place because the Bernoulli mechanism competes with the wall-wake bubble interaction, producing a periodic collision regime. \\

\begin{figure*}[htbp]
\begin{center}
\resizebox{0.49\textwidth}{!}{\includegraphics{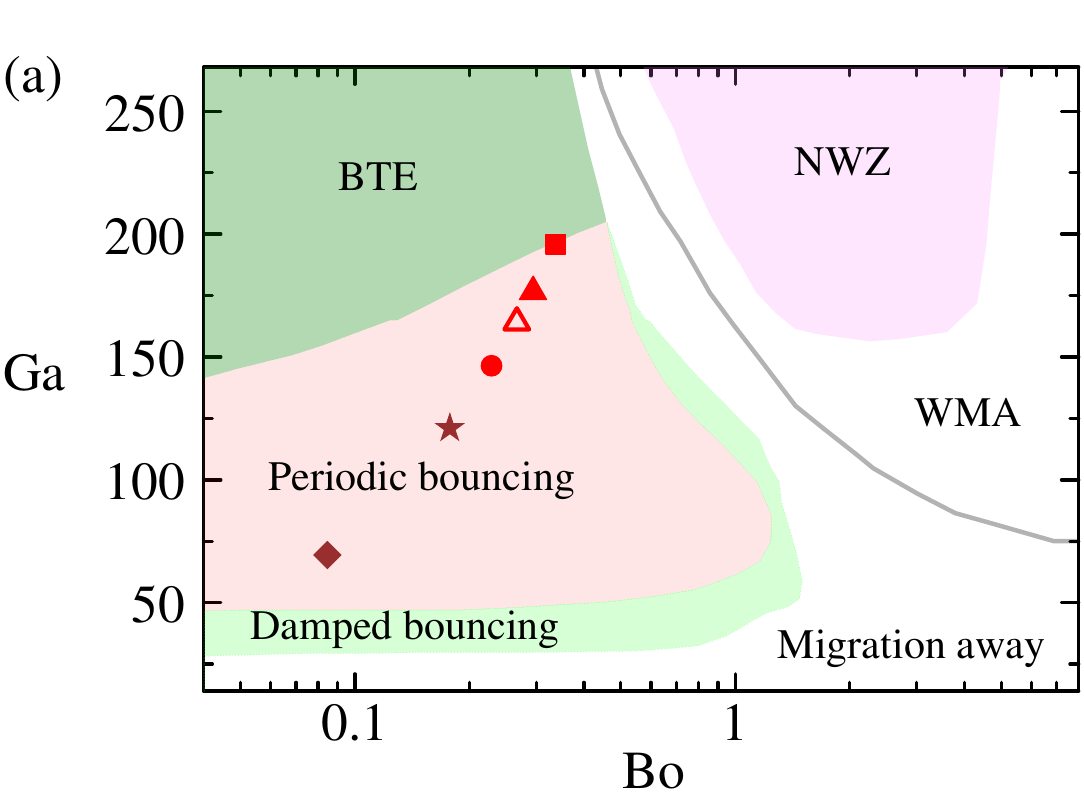}}
\resizebox{0.49\textwidth}{!}{\includegraphics{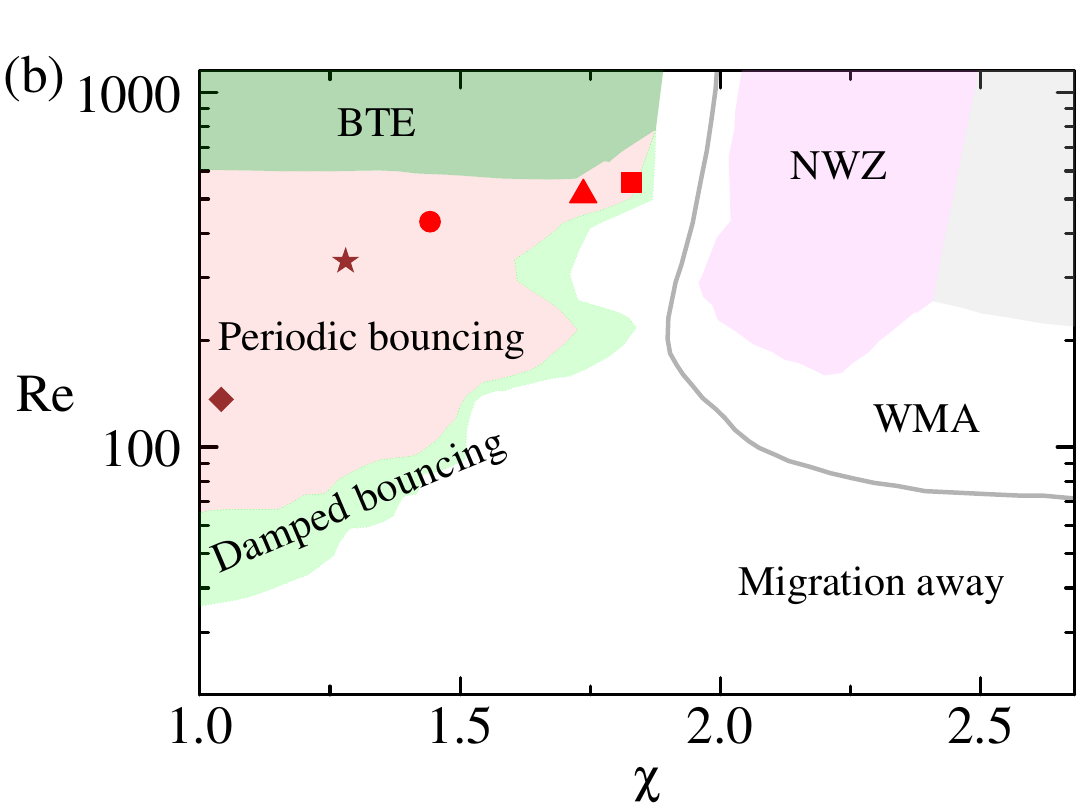}}
\end{center}
\caption{Complete state diagram (a) $Bo$-$Ga$ and (b) $\chi$-$Re$ (for $L$=1), defined here in terms of $D^{*}$, of near-wall rising regimes observed in the simulations~\citep{SZM25}. Neutral curve corresponding to the onset of path instability in a unbounded fluid (solid line)~\citep{SZM25}. The white zone straddling the neutral curve represents the whole set of conditions under which bubbles migrate away from the wall, either in the presence or in the absence of path instability. NWZ region means near-wall zigzagging, WMA wavy migration away, and BTE bouncing-tumbing-escaping. Red and brown symbols correspond to our experiments at $L\approx1$, for C+MA (collision and migration away) and PC (periodic collisions) behaviours, respectively ($D^{*}=0.79$, $D^{*}=1.15$, $D^{*}=1.28$, $D^{*}=1.48$, and $D^{*}=1.58$ mm correspond to diamond, star, circle, triangle, and square, respectively). The empty triangle correspond to a C+MA case from \citet{VBW02}, also for $L\approx1$.}
\label{mapSZM}
\end{figure*}

Finally, our experimental results are compared with recent numerical simulations reported by \citet{SZM24,SZM25} for $L=1$. Figure~\ref{mapSZM} reproduces the full state diagrams presented in \citet{SZM25}: panel (a) shows the $(Bo, Ga)$ regime map, and panel (b) displays the corresponding $(\chi, Re)$ diagram, both constructed for a fixed wall separation $L = 1$ and with dimensional quantities defined in terms of the equivalent bubble diameter $D^*$. Red and brown symbols indicate our experimental data at $L \approx 1$, corresponding to the C+MA (single collision followed by migration away) and PC (periodic collision) behaviours, respectively. For reference, a red open triangle indicates a C+MA case reported by \citet{VBW02} under comparable wall separation. Notice that all experimental data points lie within the PC region identified numerically, which includes both repeated near-wall bounces and trajectories involving a single rebound followed by escape. The brown markers, associated with PC cases, align well with the region defined by \citet{SZM25}, confirming the robustness of the regime classification. \\

In contrast, the red markers, representing C+MA behaviour, deviate qualitatively from the typical PC trajectory: after a single collision, the bubble migrates irreversibly away from the wall. This feature closely resembles the so-called BTE (bouncing–tumbling–escaping) regime introduced by \citet{SZM25}, although it arises here under different flow conditions. The observed discrepancy may be attributed to differences in the initial and boundary conditions between simulations and experiments. In the numerical setup, the wall is present from the onset ($z_w^* = 0$ mm), meaning that the bubble is continuously influenced by the wall during its entire rise, and the wake asymmetry develops from the beginning. In contrast, in our experiments, the vertical wall is located downstream ($z_w^* = 144$ mm), allowing the bubble to rise freely and develop an axisymmetric wake before encountering the wall. Upon contact, the wake must then reorganise under the influence of the wall-induced shear, which could promote the onset of wall-normal migration not observed in the simulations. A numerical investigation under the exact conditions of the experiment would be required to resolve this discrepancy, but such a comparison lies beyond the scope of the present study. Nevertheless, future work combining simulations with a controlled experimental initialisation may shed light on the transient wake reconfiguration and its role in triggering escape-like dynamics near vertical boundaries.

\section{Concluding Remarks}
\label{sec4}
A series of controlled experiments were carried out to investigate how a vertical wall influences the dynamics of deformable nitrogen bubbles rising in ultrapure water. Bubbles were generated using precision nozzles placed at the bottom of a transparent tank, and the wall was positioned at varying horizontal distances $L$ from the injector, in the range $0.3 \lesssim L \lesssim 5$. All the investigated cases exibit a rectilinear path when bubbles freely rise. Importantly, the wall did not extend to the injection point, allowing the bubbles to rise freely and reach their terminal velocity before interacting with the solid boundary.\\

High-speed imaging from two orthogonal vertical planes, enabled by a mirror system, allowed for accurate reconstruction of the bubble centroid trajectory, instantaneous velocity, and deformation, the latter quantified via the aspect ratio $\chi$. The experiments revealed four distinct regimes of bubble-wall interaction, determined by the values of $Bo$ (or $Ga$) and $L$:

\begin{enumerate}
    \item Rectilinear Path (RP): at low Bond numbers ($Bo \lesssim 0.2$) and large $L$, bubbles followed a vertical, undisturbed trajectory. The wall had negligible influence on the bubble path in this regime.

    \item Periodic Collision (PC): as $L$ decreased in the same $Bo$ range, the bubble trajectory changed significantly, entering a regime characterised by periodic lateral motion and repeated impacts with the wall. The presence of the wall perturbs the symmetry of the flow around the bubble, inducing the transition from RP to PC.

    \item Migration Away (MA): at higher Bond numbers ($Bo \gtrsim 0.2$), and for moderate to large $L$, bubbles no longer rose vertically but migrated laterally away from the wall without making contact. This behaviour results from asymmetric pressure and vorticity distributions due to bubble deformability and inertial effects.

    \item Collision and Migration Away (C+MA): when $L$ is sufficiently small at high $Bo$, bubbles initially collide with the wall and then migrate away along an inclined path. This regime combines aspects of both PC and MA.
\end{enumerate}

The systematic variation of $L$ enabled the identification of clear transitions between these regimes. It was specifically observed that the wall induces changes in the bubble trajectory, which promotes transitions between distinct regimes. These transitions are captured in a detailed regime map constructed, for the first time, in the $(Bo, L)$ parameter space. Overall, the results are in good agreement with prior experimental observations in the literature. However, they exhibit discrepancies with some recent numerical simulations—particularly concerning the onset of contact in the PC and C+MA regimes, highlighting the importance of experimental data for validating and refining computational models.\\

This study demonstrates that even in the absence of initial boundary proximity, the presence of a wall can significantly alter bubble dynamics once interaction occurs. These findings are relevant to a wide range of real-world systems, such as bubble columns, aeration tanks, and multiphase reactors, where such bubble–boundary interactions are frequent and can influence performance-critical phenomena such as mixing, mass transfer, and residence time.

\vspace{1cm}

Support from the Spanish Ministry of Science and Innovation (MCIN) through grant PID2022-140951OB-C22/ AEI/10.13039/501100011033/ FEDER, UE is gratefully acknowledged. R. Bola\~nos-Jim\'enez would like to acknowledge the University of Jaén project M.2 PDC\_1484\_UJA23 co-funded by Consejería de Universidad, Investigación e Innovación (Junta de Andalucía) and by the European Union through the FEDER Andalucía 2021–2027 Programme.

\bibliography{central}

\begin{thebibliography}{35}%
\makeatletter
\providecommand \@ifxundefined [1]{%
 \@ifx{#1\undefined}
}%
\providecommand \@ifnum [1]{%
 \ifnum #1\expandafter \@firstoftwo
 \else \expandafter \@secondoftwo
 \fi
}%
\providecommand \@ifx [1]{%
 \ifx #1\expandafter \@firstoftwo
 \else \expandafter \@secondoftwo
 \fi
}%
\providecommand \natexlab [1]{#1}%
\providecommand \enquote  [1]{``#1''}%
\providecommand \bibnamefont  [1]{#1}%
\providecommand \bibfnamefont [1]{#1}%
\providecommand \citenamefont [1]{#1}%
\providecommand \href@noop [0]{\@secondoftwo}%
\providecommand \href [0]{\begingroup \@sanitize@url \@href}%
\providecommand \@href[1]{\@@startlink{#1}\@@href}%
\providecommand \@@href[1]{\endgroup#1\@@endlink}%
\providecommand \@sanitize@url [0]{\catcode `\\12\catcode `\$12\catcode
  `\&12\catcode `\#12\catcode `\^12\catcode `\_12\catcode `\%12\relax}%
\providecommand \@@startlink[1]{}%
\providecommand \@@endlink[0]{}%
\providecommand \url  [0]{\begingroup\@sanitize@url \@url }%
\providecommand \@url [1]{\endgroup\@href {#1}{\urlprefix }}%
\providecommand \urlprefix  [0]{URL }%
\providecommand \Eprint [0]{\href }%
\providecommand \doibase [0]{http://dx.doi.org/}%
\providecommand \selectlanguage [0]{\@gobble}%
\providecommand \bibinfo  [0]{\@secondoftwo}%
\providecommand \bibfield  [0]{\@secondoftwo}%
\providecommand \translation [1]{[#1]}%
\providecommand \BibitemOpen [0]{}%
\providecommand \bibitemStop [0]{}%
\providecommand \bibitemNoStop [0]{.\EOS\space}%
\providecommand \EOS [0]{\spacefactor3000\relax}%
\providecommand \BibitemShut  [1]{\csname bibitem#1\endcsname}%
\let\auto@bib@innerbib\@empty
%</preamble>
\bibitem [{\citenamefont {Rodr\'{\i}guez-Rodr\'{\i}guez}\ \emph
  {et~al.}(2015)\citenamefont {Rodr\'{\i}guez-Rodr\'{\i}guez}, \citenamefont
  {Sevilla}, \citenamefont {Mart\'{\i}nez-Baz\'an},\ and\ \citenamefont
  {Gordillo}}]{RSMG15}%
  \BibitemOpen
  \bibfield  {author} {\bibinfo {author} {\bibfnamefont {J.}~\bibnamefont
  {Rodr\'{\i}guez-Rodr\'{\i}guez}}, \bibinfo {author} {\bibfnamefont
  {A.}~\bibnamefont {Sevilla}}, \bibinfo {author} {\bibfnamefont
  {C.}~\bibnamefont {Mart\'{\i}nez-Baz\'an}}, \ and\ \bibinfo {author}
  {\bibfnamefont {J.~M.}\ \bibnamefont {Gordillo}},\ }\bibfield  {title}
  {\enquote {\bibinfo {title} {Generation of microbubbles with applications to
  industry and medicine},}\ }\href@noop {} {\bibfield  {journal} {\bibinfo
  {journal} {Annu. Rev. Fluid Mech.}\ }\textbf {\bibinfo {volume} {47}},\
  \bibinfo {pages} {405--429} (\bibinfo {year} {2015})}\BibitemShut {NoStop}%
\bibitem [{\citenamefont {Wang}\ \emph {et~al.}(2018)\citenamefont {Wang},
  \citenamefont {Zhang}, \citenamefont {Liu}, \citenamefont {Zhang},\ and\
  \citenamefont {Cui}}]{WZLZC18}%
  \BibitemOpen
  \bibfield  {author} {\bibinfo {author} {\bibfnamefont {S.~P.}\ \bibnamefont
  {Wang}}, \bibinfo {author} {\bibfnamefont {A.~M.}\ \bibnamefont {Zhang}},
  \bibinfo {author} {\bibfnamefont {Y.~L.}\ \bibnamefont {Liu}}, \bibinfo
  {author} {\bibfnamefont {S.}~\bibnamefont {Zhang}}, \ and\ \bibinfo {author}
  {\bibfnamefont {P.}~\bibnamefont {Cui}},\ }\bibfield  {title} {\enquote
  {\bibinfo {title} {Bubble dynamics and its applications},}\ }\href@noop {}
  {\bibfield  {journal} {\bibinfo  {journal} {J. Hydrodynam.}\ }\textbf
  {\bibinfo {volume} {30}},\ \bibinfo {pages} {975--991} (\bibinfo {year}
  {2018})}\BibitemShut {NoStop}%
\bibitem [{\citenamefont {Cioncolini}\ and\ \citenamefont
  {Magnini}(2025)}]{CM25}%
  \BibitemOpen
  \bibfield  {author} {\bibinfo {author} {\bibfnamefont {A.}~\bibnamefont
  {Cioncolini}}\ and\ \bibinfo {author} {\bibfnamefont {M.}~\bibnamefont
  {Magnini}},\ }\bibfield  {title} {\enquote {\bibinfo {title} {Solitary
  bubbles rising in quiescent liquids: A critical assessment of experimental
  data and high-fidelity numerical simulations, and performance evaluation of
  selected prediction methods},}\ }\href@noop {} {\bibfield  {journal}
  {\bibinfo  {journal} {Phys. Fluids}\ }\textbf {\bibinfo {volume} {37}}
  (\bibinfo {year} {2025})}\BibitemShut {NoStop}%
\bibitem [{\citenamefont {Clift}\ \emph {et~al.}(1978)\citenamefont {Clift},
  \citenamefont {Grace},\ and\ \citenamefont {Weber}}]{CGW78}%
  \BibitemOpen
  \bibfield  {author} {\bibinfo {author} {\bibfnamefont {R.}~\bibnamefont
  {Clift}}, \bibinfo {author} {\bibfnamefont {J.~R.}\ \bibnamefont {Grace}}, \
  and\ \bibinfo {author} {\bibfnamefont {M.~E.}\ \bibnamefont {Weber}},\
  }\href@noop {} {\emph {\bibinfo {title} {Bubbles, Drops and Particles}}}\
  (\bibinfo  {publisher} {Academic Press},\ \bibinfo {address} {USA},\ \bibinfo
  {year} {1978})\BibitemShut {NoStop}%
\bibitem [{\citenamefont {Bhaga}\ and\ \citenamefont {Weber}(1981)}]{BW81}%
  \BibitemOpen
  \bibfield  {author} {\bibinfo {author} {\bibfnamefont {D.}~\bibnamefont
  {Bhaga}}\ and\ \bibinfo {author} {\bibfnamefont {M.~E.}\ \bibnamefont
  {Weber}},\ }\bibfield  {title} {\enquote {\bibinfo {title} {Bubbles in
  viscous liquids: shapes, wakes and velocities},}\ }\href@noop {} {\bibfield
  {journal} {\bibinfo  {journal} {J. Fluid Mech.}\ }\textbf {\bibinfo {volume}
  {105}},\ \bibinfo {pages} {61--85} (\bibinfo {year} {1981})}\BibitemShut
  {NoStop}%
\bibitem [{\citenamefont {Duineveld}(1995)}]{D95}%
  \BibitemOpen
  \bibfield  {author} {\bibinfo {author} {\bibfnamefont {P.C.}\ \bibnamefont
  {Duineveld}},\ }\bibfield  {title} {\enquote {\bibinfo {title} {The rise
  velocity and shape of bubbles in pure water at high reynolds number},}\
  }\href@noop {} {\bibfield  {journal} {\bibinfo  {journal} {J. Fluid Mech.}\
  }\textbf {\bibinfo {volume} {292}},\ \bibinfo {pages} {325–332} (\bibinfo
  {year} {1995})}\BibitemShut {NoStop}%
\bibitem [{\citenamefont {Magnaudet}\ and\ \citenamefont {Eames}(2000)}]{ME00}%
  \BibitemOpen
  \bibfield  {author} {\bibinfo {author} {\bibfnamefont {J.}~\bibnamefont
  {Magnaudet}}\ and\ \bibinfo {author} {\bibfnamefont {I.}~\bibnamefont
  {Eames}},\ }\bibfield  {title} {\enquote {\bibinfo {title} {The motion of
  high-reynolds-number bubbles in inhomogeneous flows},}\ }\href@noop {}
  {\bibfield  {journal} {\bibinfo  {journal} {Annu. Rev. Fluid Mech.}\ }\textbf
  {\bibinfo {volume} {32}},\ \bibinfo {pages} {659--708} (\bibinfo {year}
  {2000})}\BibitemShut {NoStop}%
\bibitem [{\citenamefont {Cano-Lozano}\ \emph
  {et~al.}(2016{\natexlab{a}})\citenamefont {Cano-Lozano}, \citenamefont
  {Mart\'{\i}nez-Baz\'an}, \citenamefont {Magnaudet},\ and\ \citenamefont
  {Tchoufag}}]{CMMT16}%
  \BibitemOpen
  \bibfield  {author} {\bibinfo {author} {\bibfnamefont {J.~C.}\ \bibnamefont
  {Cano-Lozano}}, \bibinfo {author} {\bibfnamefont {C.}~\bibnamefont
  {Mart\'{\i}nez-Baz\'an}}, \bibinfo {author} {\bibfnamefont {J.}~\bibnamefont
  {Magnaudet}}, \ and\ \bibinfo {author} {\bibfnamefont {J.}~\bibnamefont
  {Tchoufag}},\ }\bibfield  {title} {\enquote {\bibinfo {title} {Paths and
  wakes of deformable nearly spheroidal rising bubbles close to the transition
  to path instability},}\ }\href@noop {} {\bibfield  {journal} {\bibinfo
  {journal} {Phys. Rev. Fluids}\ }\textbf {\bibinfo {volume} {1}},\ \bibinfo
  {pages} {053604} (\bibinfo {year} {2016}{\natexlab{a}})}\BibitemShut
  {NoStop}%
\bibitem [{\citenamefont {Bonnefis}\ \emph {et~al.}(2023)\citenamefont
  {Bonnefis}, \citenamefont {Fabre},\ and\ \citenamefont {Magnaudet}}]{BFM23}%
  \BibitemOpen
  \bibfield  {author} {\bibinfo {author} {\bibfnamefont {P.}~\bibnamefont
  {Bonnefis}}, \bibinfo {author} {\bibfnamefont {D.}~\bibnamefont {Fabre}}, \
  and\ \bibinfo {author} {\bibfnamefont {J.}~\bibnamefont {Magnaudet}},\
  }\bibfield  {title} {\enquote {\bibinfo {title} {When, how, and why the path
  of an air bubble rising in pure water becomes unstable},}\ }\href@noop {}
  {\bibfield  {journal} {\bibinfo  {journal} {Proc. Natl. Acad. Sci.}\ }\textbf
  {\bibinfo {volume} {120}},\ \bibinfo {pages} {e2300897120} (\bibinfo {year}
  {2023})}\BibitemShut {NoStop}%
\bibitem [{\citenamefont {Rubio}\ \emph {et~al.}(2024)\citenamefont {Rubio},
  \citenamefont {Vega}, \citenamefont {Cabezas}, \citenamefont {Montanero},
  \citenamefont {López-Herrera},\ and\ \citenamefont {Herrada}}]{RVCMLH24}%
  \BibitemOpen
  \bibfield  {author} {\bibinfo {author} {\bibfnamefont {A.}~\bibnamefont
  {Rubio}}, \bibinfo {author} {\bibfnamefont {E.J.}\ \bibnamefont {Vega}},
  \bibinfo {author} {\bibfnamefont {M.G.}\ \bibnamefont {Cabezas}}, \bibinfo
  {author} {\bibfnamefont {J.~M.}\ \bibnamefont {Montanero}}, \bibinfo {author}
  {\bibfnamefont {J.~M.}\ \bibnamefont {López-Herrera}}, \ and\ \bibinfo
  {author} {\bibfnamefont {M.~A.}\ \bibnamefont {Herrada}},\ }\bibfield
  {title} {\enquote {\bibinfo {title} {Bubble rising in the presence of a
  surfactant at very low concentrations},}\ }\href@noop {} {\bibfield
  {journal} {\bibinfo  {journal} {Physics of Fluids}\ }\textbf {\bibinfo
  {volume} {36}},\ \bibinfo {pages} {062112} (\bibinfo {year}
  {2024})}\BibitemShut {NoStop}%
\bibitem [{\citenamefont {Fern\'andez-Mart\'{\i}nez}\ \emph
  {et~al.}(2025)\citenamefont {Fern\'andez-Mart\'{\i}nez}, \citenamefont
  {Cabezas}, \citenamefont {L\'opez-Herrera}, \citenamefont {Herrada},\ and\
  \citenamefont {Montanero}}]{FCLHM25}%
  \BibitemOpen
  \bibfield  {author} {\bibinfo {author} {\bibfnamefont {D.}~\bibnamefont
  {Fern\'andez-Mart\'{\i}nez}}, \bibinfo {author} {\bibfnamefont {M.~G.}\
  \bibnamefont {Cabezas}}, \bibinfo {author} {\bibfnamefont {J.~M.}\
  \bibnamefont {L\'opez-Herrera}}, \bibinfo {author} {\bibfnamefont {M.~A.}\
  \bibnamefont {Herrada}}, \ and\ \bibinfo {author} {\bibfnamefont {J.~M.}\
  \bibnamefont {Montanero}},\ }\bibfield  {title} {\enquote {\bibinfo {title}
  {Transient bubble rising in the presence of a surfactant at very low
  concentrations},}\ }\href@noop {} {\bibfield  {journal} {\bibinfo  {journal}
  {Int. J. Multiphase Flow}\ }\textbf {\bibinfo {volume} {188}},\ \bibinfo
  {pages} {105205} (\bibinfo {year} {2025})}\BibitemShut {NoStop}%
\bibitem [{\citenamefont {Blanco}\ and\ \citenamefont
  {Magnaudet}(1995)}]{BM95}%
  \BibitemOpen
  \bibfield  {author} {\bibinfo {author} {\bibfnamefont {A.}~\bibnamefont
  {Blanco}}\ and\ \bibinfo {author} {\bibfnamefont {J.}~\bibnamefont
  {Magnaudet}},\ }\bibfield  {title} {\enquote {\bibinfo {title} {The structure
  of the axisymmetric high-{R}eynolds number flow around an ellipsoidal bubble
  of fixed shape},}\ }\href@noop {} {\bibfield  {journal} {\bibinfo  {journal}
  {Phys. Fluids}\ }\textbf {\bibinfo {volume} {7}},\ \bibinfo {pages}
  {1265--1274} (\bibinfo {year} {1995})}\BibitemShut {NoStop}%
\bibitem [{\citenamefont {Mougin}\ and\ \citenamefont
  {Magnaudet}(2002)}]{MM02}%
  \BibitemOpen
  \bibfield  {author} {\bibinfo {author} {\bibfnamefont {G.}~\bibnamefont
  {Mougin}}\ and\ \bibinfo {author} {\bibfnamefont {J.}~\bibnamefont
  {Magnaudet}},\ }\bibfield  {title} {\enquote {\bibinfo {title} {Path
  instability of a rising bubble},}\ }\href@noop {} {\bibfield  {journal}
  {\bibinfo  {journal} {Phys. Rev. Lett.}\ }\textbf {\bibinfo {volume} {88}},\
  \bibinfo {pages} {014502} (\bibinfo {year} {2002})}\BibitemShut {NoStop}%
\bibitem [{\citenamefont {Yang}\ \emph {et~al.}(2003)\citenamefont {Yang},
  \citenamefont {Prosperetti},\ and\ \citenamefont {Takagi}}]{YPT03}%
  \BibitemOpen
  \bibfield  {author} {\bibinfo {author} {\bibfnamefont {B.}~\bibnamefont
  {Yang}}, \bibinfo {author} {\bibfnamefont {A.}~\bibnamefont {Prosperetti}}, \
  and\ \bibinfo {author} {\bibfnamefont {S.}~\bibnamefont {Takagi}},\
  }\bibfield  {title} {\enquote {\bibinfo {title} {The transient rise of a
  bubble subject to shape or volume changes},}\ }\href@noop {} {\bibfield
  {journal} {\bibinfo  {journal} {Phys. Fluids}\ }\textbf {\bibinfo {volume}
  {15}},\ \bibinfo {pages} {2640--2648} (\bibinfo {year} {2003})}\BibitemShut
  {NoStop}%
\bibitem [{\citenamefont {Zenit}\ and\ \citenamefont {Magnaudet}(2008)}]{ZM08}%
  \BibitemOpen
  \bibfield  {author} {\bibinfo {author} {\bibfnamefont {R.}~\bibnamefont
  {Zenit}}\ and\ \bibinfo {author} {\bibfnamefont {J.}~\bibnamefont
  {Magnaudet}},\ }\bibfield  {title} {\enquote {\bibinfo {title} {Path
  instability of rising spheroidal air bubbles: A shape-controlled process},}\
  }\href@noop {} {\bibfield  {journal} {\bibinfo  {journal} {Phys. Fluids}\
  }\textbf {\bibinfo {volume} {20}},\ \bibinfo {pages} {061702} (\bibinfo
  {year} {2008})}\BibitemShut {NoStop}%
\bibitem [{\citenamefont {Tripathi}\ \emph {et~al.}(2015)\citenamefont
  {Tripathi}, \citenamefont {Sahu},\ and\ \citenamefont
  {Govindarajan}}]{TSG15}%
  \BibitemOpen
  \bibfield  {author} {\bibinfo {author} {\bibfnamefont {M.~K.}\ \bibnamefont
  {Tripathi}}, \bibinfo {author} {\bibfnamefont {K.~C.}\ \bibnamefont {Sahu}},
  \ and\ \bibinfo {author} {\bibfnamefont {R.}~\bibnamefont {Govindarajan}},\
  }\bibfield  {title} {\enquote {\bibinfo {title} {Dynamics of an initially
  spherical bubble rising in quiescent liquid},}\ }\href@noop {} {\bibfield
  {journal} {\bibinfo  {journal} {Nat Commun}\ }\textbf {\bibinfo {volume}
  {6}},\ \bibinfo {pages} {6268} (\bibinfo {year} {2015})}\BibitemShut
  {NoStop}%
\bibitem [{\citenamefont {Magnaudet}(2003)}]{M03c}%
  \BibitemOpen
  \bibfield  {author} {\bibinfo {author} {\bibfnamefont {J.}~\bibnamefont
  {Magnaudet}},\ }\bibfield  {title} {\enquote {\bibinfo {title} {Small
  inertial effects on a spherical bubble, drop or particle moving near a wall
  in a time-dependent linear flow},}\ }\href@noop {} {\bibfield  {journal}
  {\bibinfo  {journal} {J. Fluid Mech.}\ }\textbf {\bibinfo {volume} {485}},\
  \bibinfo {pages} {115--142} (\bibinfo {year} {2003})}\BibitemShut {NoStop}%
\bibitem [{\citenamefont {Sugiyama}\ and\ \citenamefont
  {Takemura}(2010)}]{ST10}%
  \BibitemOpen
  \bibfield  {author} {\bibinfo {author} {\bibfnamefont {K.}~\bibnamefont
  {Sugiyama}}\ and\ \bibinfo {author} {\bibfnamefont {F.}~\bibnamefont
  {Takemura}},\ }\bibfield  {title} {\enquote {\bibinfo {title} {On the lateral
  migration of a slightly deformed bubble rising near a vertical plane wall},}\
  }\href@noop {} {\bibfield  {journal} {\bibinfo  {journal} {J. Fluid Mech.}\
  }\textbf {\bibinfo {volume} {662}},\ \bibinfo {pages} {209--231} (\bibinfo
  {year} {2010})}\BibitemShut {NoStop}%
\bibitem [{\citenamefont {Zhang}\ \emph {et~al.}(2020)\citenamefont {Zhang},
  \citenamefont {Dabiri}, \citenamefont {Chen},\ and\ \citenamefont
  {You}}]{ZDCY20}%
  \BibitemOpen
  \bibfield  {author} {\bibinfo {author} {\bibfnamefont {Y.}~\bibnamefont
  {Zhang}}, \bibinfo {author} {\bibfnamefont {S.}~\bibnamefont {Dabiri}},
  \bibinfo {author} {\bibfnamefont {K.}~\bibnamefont {Chen}}, \ and\ \bibinfo
  {author} {\bibfnamefont {Y.}~\bibnamefont {You}},\ }\bibfield  {title}
  {\enquote {\bibinfo {title} {An initially spherical bubble rising near a
  vertical wall},}\ }\href@noop {} {\bibfield  {journal} {\bibinfo  {journal}
  {Int. J. Heat Fluid Flow}\ }\textbf {\bibinfo {volume} {85}},\ \bibinfo
  {pages} {108649} (\bibinfo {year} {2020})}\BibitemShut {NoStop}%
\bibitem [{\citenamefont {Shi}\ \emph {et~al.}(2024)\citenamefont {Shi},
  \citenamefont {Zhang},\ and\ \citenamefont {Magnaudet}}]{SZM24}%
  \BibitemOpen
  \bibfield  {author} {\bibinfo {author} {\bibfnamefont {P.}~\bibnamefont
  {Shi}}, \bibinfo {author} {\bibfnamefont {J.}~\bibnamefont {Zhang}}, \ and\
  \bibinfo {author} {\bibfnamefont {J.}~\bibnamefont {Magnaudet}},\ }\bibfield
  {title} {\enquote {\bibinfo {title} {Lateral migration and bouncing of a
  deformable bubble rising near a vertical wall. part 1. moderately inertial
  regimes},}\ }\href@noop {} {\bibfield  {journal} {\bibinfo  {journal} {J.
  Fluid Mech.}\ }\textbf {\bibinfo {volume} {998}},\ \bibinfo {pages} {A8}
  (\bibinfo {year} {2024})}\BibitemShut {NoStop}%
\bibitem [{\citenamefont {Shi}(2024)}]{Shi24}%
  \BibitemOpen
  \bibfield  {author} {\bibinfo {author} {\bibfnamefont {P.}~\bibnamefont
  {Shi}},\ }\bibfield  {title} {\enquote {\bibinfo {title} {Reversal of the
  transverse force on a spherical bubble rising close to a vertical wall at
  moderate-to-high reynolds numbers},}\ }\href@noop {} {\bibfield  {journal}
  {\bibinfo  {journal} {Phys. Rev. Fluids}\ }\textbf {\bibinfo {volume} {9}},\
  \bibinfo {pages} {023601} (\bibinfo {year} {2024})}\BibitemShut {NoStop}%
\bibitem [{\citenamefont {Takemura}\ \emph {et~al.}(2002)\citenamefont
  {Takemura}, \citenamefont {Takagi},\ and\ \citenamefont
  {Matsumoto}}]{TTMM02}%
  \BibitemOpen
  \bibfield  {author} {\bibinfo {author} {\bibfnamefont {F.}~\bibnamefont
  {Takemura}}, \bibinfo {author} {\bibfnamefont {S.}~\bibnamefont {Takagi}}, \
  and\ \bibinfo {author} {\bibfnamefont {J.~Magnaudet.~Y.}\ \bibnamefont
  {Matsumoto}},\ }\bibfield  {title} {\enquote {\bibinfo {title} {Drag and lift
  forces on a bubble rising near a vertical wall in a viscous liquid},}\
  }\href@noop {} {\bibfield  {journal} {\bibinfo  {journal} {J. Fluid Mech.}\
  }\textbf {\bibinfo {volume} {461}},\ \bibinfo {pages} {277--300} (\bibinfo
  {year} {2002})}\BibitemShut {NoStop}%
\bibitem [{\citenamefont {Takemura}\ and\ \citenamefont
  {Magnaudet}(2003)}]{TM03}%
  \BibitemOpen
  \bibfield  {author} {\bibinfo {author} {\bibfnamefont {F.}~\bibnamefont
  {Takemura}}\ and\ \bibinfo {author} {\bibfnamefont {J.}~\bibnamefont
  {Magnaudet}},\ }\bibfield  {title} {\enquote {\bibinfo {title} {The
  transverse force on clean and contaminated bubbles rising near a vertical
  wall at moderate reynolds number},}\ }\href@noop {} {\bibfield  {journal}
  {\bibinfo  {journal} {J. Fluid Mech.}\ }\textbf {\bibinfo {volume} {495}},\
  \bibinfo {pages} {235--253} (\bibinfo {year} {2003})}\BibitemShut {NoStop}%
\bibitem [{\citenamefont {Sugioka}\ and\ \citenamefont {Tsukada}(2015)}]{ST15}%
  \BibitemOpen
  \bibfield  {author} {\bibinfo {author} {\bibfnamefont {K.}~\bibnamefont
  {Sugioka}}\ and\ \bibinfo {author} {\bibfnamefont {T.}~\bibnamefont
  {Tsukada}},\ }\bibfield  {title} {\enquote {\bibinfo {title} {Direct
  numerical simulations of drag and lift forces acting on a spherical bubble
  near a plane wall},}\ }\href@noop {} {\bibfield  {journal} {\bibinfo
  {journal} {Int. J. Multiph. Flow}\ }\textbf {\bibinfo {volume} {71}},\
  \bibinfo {pages} {32--37} (\bibinfo {year} {2015})}\BibitemShut {NoStop}%
\bibitem [{\citenamefont {Shi}\ \emph {et~al.}(2025)\citenamefont {Shi},
  \citenamefont {Zhang},\ and\ \citenamefont {Magnaudet}}]{SZM25}%
  \BibitemOpen
  \bibfield  {author} {\bibinfo {author} {\bibfnamefont {P.}~\bibnamefont
  {Shi}}, \bibinfo {author} {\bibfnamefont {J.}~\bibnamefont {Zhang}}, \ and\
  \bibinfo {author} {\bibfnamefont {J.}~\bibnamefont {Magnaudet}},\ }\bibfield
  {title} {\enquote {\bibinfo {title} {Lateral migration and bouncing of a
  deformable bubble rising near a vertical wall. part 2. highly inertial
  regimes},}\ }\href@noop {} {\bibfield  {journal} {\bibinfo  {journal} {J.
  Fluid Mech,}\ }\textbf {\bibinfo {volume} {1013}},\ \bibinfo {pages} {A19}
  (\bibinfo {year} {2025})}\BibitemShut {NoStop}%
\bibitem [{\citenamefont {{de Vries}}\ \emph {et~al.}(2002)\citenamefont {{de
  Vries}}, \citenamefont {Biesheuvel},\ and\ \citenamefont {{van
  Wijngaarden}}}]{VBW02}%
  \BibitemOpen
  \bibfield  {author} {\bibinfo {author} {\bibfnamefont {A.W.G}\ \bibnamefont
  {{de Vries}}}, \bibinfo {author} {\bibfnamefont {A.}~\bibnamefont
  {Biesheuvel}}, \ and\ \bibinfo {author} {\bibfnamefont {L.}~\bibnamefont
  {{van Wijngaarden}}},\ }\bibfield  {title} {\enquote {\bibinfo {title} {Notes
  on the path and wake of a gas bubble rising in pure water},}\ }\href@noop {}
  {\bibfield  {journal} {\bibinfo  {journal} {Int. J. Multiphase Flow}\
  }\textbf {\bibinfo {volume} {28}},\ \bibinfo {pages} {1823--1835} (\bibinfo
  {year} {2002})}\BibitemShut {NoStop}%
\bibitem [{\citenamefont {Jeong}\ and\ \citenamefont {Park}(2015)}]{JP15}%
  \BibitemOpen
  \bibfield  {author} {\bibinfo {author} {\bibfnamefont {H.}~\bibnamefont
  {Jeong}}\ and\ \bibinfo {author} {\bibfnamefont {H.}~\bibnamefont {Park}},\
  }\bibfield  {title} {\enquote {\bibinfo {title} {Near-wall rising behaviour
  of a deformable bubble at high {R}eynolds number},}\ }\href@noop {}
  {\bibfield  {journal} {\bibinfo  {journal} {J. Fluid Mech.}\ }\textbf
  {\bibinfo {volume} {771}},\ \bibinfo {pages} {564--594} (\bibinfo {year}
  {2015})}\BibitemShut {NoStop}%
\bibitem [{\citenamefont {Joohyoung}\ and\ \citenamefont
  {Hyungmin}(2017)}]{JH17}%
  \BibitemOpen
  \bibfield  {author} {\bibinfo {author} {\bibfnamefont {L.}~\bibnamefont
  {Joohyoung}}\ and\ \bibinfo {author} {\bibfnamefont {P.}~\bibnamefont
  {Hyungmin}},\ }\bibfield  {title} {\enquote {\bibinfo {title} {Wake
  structures behind an oscillating bubble rising close to a vertical wall},}\
  }\href@noop {} {\bibfield  {journal} {\bibinfo  {journal} {Int. J. Multiph.
  Flow}\ }\textbf {\bibinfo {volume} {91}},\ \bibinfo {pages} {225--242}
  (\bibinfo {year} {2017})}\BibitemShut {NoStop}%
\bibitem [{\citenamefont {Estepa-Cantero}\ \emph {et~al.}(2024)\citenamefont
  {Estepa-Cantero}, \citenamefont {Mart{\'\i}nez-Baz{\'a}n},\ and\
  \citenamefont {Bola{\~n}os-Jim{\'e}nez}}]{ECMBBJ24}%
  \BibitemOpen
  \bibfield  {author} {\bibinfo {author} {\bibfnamefont {C.}~\bibnamefont
  {Estepa-Cantero}}, \bibinfo {author} {\bibfnamefont {C.}~\bibnamefont
  {Mart{\'\i}nez-Baz{\'a}n}}, \ and\ \bibinfo {author} {\bibfnamefont
  {R.}~\bibnamefont {Bola{\~n}os-Jim{\'e}nez}},\ }\bibfield  {title} {\enquote
  {\bibinfo {title} {Bubble rising near a vertical wall: Experimental
  characterization of paths and velocity},}\ }\href@noop {} {\bibfield
  {journal} {\bibinfo  {journal} {Phys. Fluids}\ }\textbf {\bibinfo {volume}
  {36}} (\bibinfo {year} {2024})}\BibitemShut {NoStop}%
\bibitem [{\citenamefont {Ferrera}\ \emph {et~al.}(2007)\citenamefont
  {Ferrera}, \citenamefont {Montanero},\ and\ \citenamefont {Cabezas}}]{FMC07}%
  \BibitemOpen
  \bibfield  {author} {\bibinfo {author} {\bibfnamefont {C.}~\bibnamefont
  {Ferrera}}, \bibinfo {author} {\bibfnamefont {J.~M.}\ \bibnamefont
  {Montanero}}, \ and\ \bibinfo {author} {\bibfnamefont {M.~G.}\ \bibnamefont
  {Cabezas}},\ }\bibfield  {title} {\enquote {\bibinfo {title} {An analysis of
  the sensitivity of pendant drops and liquid bridges to measure the
  interfacial tension},}\ }\href@noop {} {\bibfield  {journal} {\bibinfo
  {journal} {Meas. Sci. Technol.}\ }\textbf {\bibinfo {volume} {18}},\ \bibinfo
  {pages} {3713--3723} (\bibinfo {year} {2007})}\BibitemShut {NoStop}%
\bibitem [{\citenamefont {Luo}\ \emph {et~al.}(2022)\citenamefont {Luo},
  \citenamefont {Wang}, \citenamefont {Zhang}, \citenamefont {Guo},
  \citenamefont {Zheng}, \citenamefont {Xiang}, \citenamefont {Liu},\ and\
  \citenamefont {Liu}}]{LWZGZXLL22}%
  \BibitemOpen
  \bibfield  {author} {\bibinfo {author} {\bibfnamefont {Y.}~\bibnamefont
  {Luo}}, \bibinfo {author} {\bibfnamefont {Z.}~\bibnamefont {Wang}}, \bibinfo
  {author} {\bibfnamefont {B.}~\bibnamefont {Zhang}}, \bibinfo {author}
  {\bibfnamefont {K.}~\bibnamefont {Guo}}, \bibinfo {author} {\bibfnamefont
  {L.}~\bibnamefont {Zheng}}, \bibinfo {author} {\bibfnamefont
  {W.}~\bibnamefont {Xiang}}, \bibinfo {author} {\bibfnamefont
  {H.}~\bibnamefont {Liu}}, \ and\ \bibinfo {author} {\bibfnamefont
  {C.}~\bibnamefont {Liu}},\ }\bibfield  {title} {\enquote {\bibinfo {title}
  {Experimental study of the effect of the surfactant on the single bubble
  rising in stagnant surfactant solutions and a mathematical model for the
  bubble motion},}\ }\href@noop {} {\bibfield  {journal} {\bibinfo  {journal}
  {Ind. Eng. Chem. Res.}\ }\textbf {\bibinfo {volume} {61}},\ \bibinfo {pages}
  {9514--9527} (\bibinfo {year} {2022})}\BibitemShut {NoStop}%
\bibitem [{\citenamefont {Herrada}\ and\ \citenamefont {Eggers}(2023)}]{HE23}%
  \BibitemOpen
  \bibfield  {author} {\bibinfo {author} {\bibfnamefont {M.A.}\ \bibnamefont
  {Herrada}}\ and\ \bibinfo {author} {\bibfnamefont {J.G.}\ \bibnamefont
  {Eggers}},\ }\bibfield  {title} {\enquote {\bibinfo {title} {Leonardo’s
  paradox resolved: path instability of an air bubble rising in water},}\
  }\href@noop {} {\bibfield  {journal} {\bibinfo  {journal} {Proc. Natl. Acad.
  Sci.}\ }\textbf {\bibinfo {volume} {120}},\ \bibinfo {pages} {e2216830120}
  (\bibinfo {year} {2023})}\BibitemShut {NoStop}%
\bibitem [{\citenamefont {Cano-Lozano}\ \emph
  {et~al.}(2016{\natexlab{b}})\citenamefont {Cano-Lozano}, \citenamefont
  {Tchoufag}, \citenamefont {Magnaudet},\ and\ \citenamefont
  {Mart\'{\i}nez-Baz\'an}}]{CTMM16}%
  \BibitemOpen
  \bibfield  {author} {\bibinfo {author} {\bibfnamefont {J.C.}\ \bibnamefont
  {Cano-Lozano}}, \bibinfo {author} {\bibfnamefont {J.}~\bibnamefont
  {Tchoufag}}, \bibinfo {author} {\bibfnamefont {J.}~\bibnamefont {Magnaudet}},
  \ and\ \bibinfo {author} {\bibfnamefont {C.}~\bibnamefont
  {Mart\'{\i}nez-Baz\'an}},\ }\bibfield  {title} {\enquote {\bibinfo {title} {A
  global stability approach to wake and path instabilities of nearly oblate
  spheroidal rising bubbles},}\ }\href@noop {} {\bibfield  {journal} {\bibinfo
  {journal} {Phys. Fluids}\ }\textbf {\bibinfo {volume} {198}},\ \bibinfo
  {pages} {014102} (\bibinfo {year} {2016}{\natexlab{b}})}\BibitemShut
  {NoStop}%
\bibitem [{\citenamefont {Yan}\ \emph {et~al.}(2022)\citenamefont {Yan},
  \citenamefont {Zhang}, \citenamefont {Liao}, \citenamefont {Zhang},
  \citenamefont {Zhou},\ and\ \citenamefont {Liu}}]{YZLZZL22}%
  \BibitemOpen
  \bibfield  {author} {\bibinfo {author} {\bibfnamefont {H.}~\bibnamefont
  {Yan}}, \bibinfo {author} {\bibfnamefont {H.}~\bibnamefont {Zhang}}, \bibinfo
  {author} {\bibfnamefont {Y.}~\bibnamefont {Liao}}, \bibinfo {author}
  {\bibfnamefont {H.}~\bibnamefont {Zhang}}, \bibinfo {author} {\bibfnamefont
  {P.}~\bibnamefont {Zhou}}, \ and\ \bibinfo {author} {\bibfnamefont
  {L.}~\bibnamefont {Liu}},\ }\bibfield  {title} {\enquote {\bibinfo {title} {A
  single bubble rising in the vicinity of a vertical wall: A numerical study
  based on volume of fluid method},}\ }\href@noop {} {\bibfield  {journal}
  {\bibinfo  {journal} {Ocean Eng.}\ }\textbf {\bibinfo {volume} {263}},\
  \bibinfo {pages} {112379} (\bibinfo {year} {2022})}\BibitemShut {NoStop}%
\bibitem [{\citenamefont {Rastello}\ \emph {et~al.}(2011)\citenamefont
  {Rastello}, \citenamefont {Mari{\'e}},\ and\ \citenamefont {Lance}}]{RML11}%
  \BibitemOpen
  \bibfield  {author} {\bibinfo {author} {\bibfnamefont {M.}~\bibnamefont
  {Rastello}}, \bibinfo {author} {\bibfnamefont {J.L.}\ \bibnamefont
  {Mari{\'e}}}, \ and\ \bibinfo {author} {\bibfnamefont {M.}~\bibnamefont
  {Lance}},\ }\bibfield  {title} {\enquote {\bibinfo {title} {Drag and lift
  forces on clean spherical and ellipsoidal bubbles in a solid-body rotating
  flow},}\ }\href@noop {} {\bibfield  {journal} {\bibinfo  {journal} {J. Fluid
  Mech.}\ }\textbf {\bibinfo {volume} {682}},\ \bibinfo {pages} {434--459}
  (\bibinfo {year} {2011})}\BibitemShut {NoStop}%
\end{thebibliography}%

\end{document}